\documentclass[prl,amsmath,amssymb,twocolumn]{revtex4-2}
\usepackage{graphicx}
\usepackage{subfigure}
\usepackage{adjustbox}
\usepackage{bm}
\usepackage{color}
\usepackage{braket}
\usepackage{standalone}
\usepackage{multirow}
\usepackage{tikz}
\usepackage{mathrsfs}
\usepackage{dsfont}

\usepackage[colorlinks,bookmarks=true,citecolor=blue,linkcolor=red,urlcolor=blue]{hyperref}

\newcommand{\be}{\begin{equation}}
\newcommand{\ee}{\end{equation}}

\newcommand{\ba}{\begin{eqnarray}}
\newcommand{\ea}{\end{eqnarray}}

\newcommand*{\id}{{\normalfont\hbox{1\kern-0.15em \vrule width .8pt depth-.5pt}}}

\begin{document}

\title{Quench dynamics of collective modes in fractional quantum Hall bilayers}
\author{Zhao Liu$^{1}$, Ajit C. Balram$^{2}$, Zlatko Papi\'c$^{3}$, and Andrey Gromov$^{4}$}
\affiliation{$^{1}$Zhejiang Institute of Modern Physics, Zhejiang University, Hangzhou 310027, China}
\affiliation{$^{2}$Institute of Mathematical Sciences, HBNI, CIT Campus, Chennai 600113, India}
\affiliation{$^{3}$School of Physics and Astronomy, University of Leeds, Leeds LS2 9JT, United Kingdom}
\affiliation{$^{4}$Brown Theoretical Physics Center and Department of Physics, Brown University, 182 Hope Street, Providence, RI 02912, USA}

\date{\today}

\begin{abstract}
We introduce different types of quenches to probe the non-equilibrium dynamics and multiple collective modes of bilayer fractional quantum Hall states. We show that applying an electric field in one layer induces oscillations of a spin-$1$ degree of freedom, whose frequency matches the long-wavelength limit of the dipole mode. On the other hand, oscillations of the long-wavelength limit of the quadrupole mode, i.e., the spin-$2$ graviton, as well as the combination of two spin-$1$ states, can be activated by a sudden change of band mass anisotropy.  We construct an effective field theory to describe the quench dynamics of these collective modes. In particular, we derive the dynamics for both the spin-$2$ and the spin-$1$ states and demonstrate their excellent agreement with numerics.

\end{abstract}

\maketitle

{\it Introduction.} A paradigmatic property of condensed phases of matter is the existence of a collective mode -- coherent oscillations of the medium -- which governs the system's low-energy physics~\cite{AndersonConcepts}. The Feynman-Bijl ansatz~\cite{FeynmanStatistical} or ``single-mode approximation'' (SMA) is an elegant formulation of this idea, originally applied to understand the emergent phonon and roton excitations in liquid helium. The same idea has found applications in correlated systems, such as the plasmon modes in three-dimensional electron systems~\cite{Lundqvist1967, Overhauser1971} and one-dimensional (1D) quantum spin systems~\cite{Affleck88, Arovas1988, Takahashi1984, Sorensen1994, Arovas1989, Thomale2015, Moudgalya2018}. Recent progress in tensor networks has enabled accurate descriptions of collective modes in both 1D and 2D lattice systems~\cite{Haegeman2012, Vanderstraeten2019}. 

Collective excitations are also ubiquitous in strongly-correlated topological phases in two-dimensional electron gases (2DEGs), which are experimentally observed in the regime of the fractional quantum Hall (FQH) effect~\cite{Tsui82}. While there has been much focus on understanding the properties of \emph{charged} excitations of FQH phases, fueled by their exotic properties such as fractional charge and fractional statistics~\cite{Laughlin83, Arovas84, Moore91}, recently there has been a resurgence of interest in the \emph{neutral} collective modes of FQH systems, some of which are also accurately described using the SMA~\cite{Girvin85, Girvin86, Renn93, Yang12b, Repellin2014}. In comparison with 1D or topologically-trivial systems, the FQH collective modes are endowed with additional physical properties, which makes their physics much richer. For example, it has recently been realized that the long-wavelength limit of the Girvin-MacDonald-Platzman (GMP) mode~\cite{Girvin85, Girvin86} exhibits an emergent quantum geometry~\cite{Haldane11, Gromov17, nguyen2018fractional}. This geometric degree of freedom has been dubbed FQH ``graviton" since it carries angular momentum $L=2$, reminiscent of the spin-2 elementary particle~\cite{Yang12b, Golkar2016, Gromov17, gromov2017investigating}. The conventional probes of FQH collective modes by inelastic light scattering~\cite{Pinczuk93, Platzman96, Kang01, Kukushkin09} are limited to finite momenta $k$,  thus they can only indirectly measure the graviton which emerges in $k\to 0$ limit. In contrast, recent works in single-layer FQH systems~\cite{Liu18, Lapa19} have shown that the graviton can be directly excited in a dynamical quench experiment, where the band mass tensor of the 2DEG is suddenly made anisotropic or the magnetic field is abruptly tilted (see also a recent proposal using surface acoustic waves~\cite{Liou19}). 


Despite this progress in understanding the dynamics of the collective mode in single-layer FQH systems, many interesting new questions arise in multicomponent FQH systems~\cite{Girvin07}, such as FQH bilayers. The additional layer degree of freedom gives rise to \emph{multiple} collective excitations~\cite{Renn93, MacDonald94, Moon95, Shizuya04}, thereby presenting a new avenue to study the non-equilibrium dynamics of FQH systems. In this Letter, we show that FQH bilayers provide a versatile platform to probe the dynamics of individual or coupled collective modes with rich topological and geometric properties. We report the investigation of an FQH bilayer system of bosons at total filling $\nu=2/3$, which hosts two collective modes: a spin-2 excitation (graviton or quadrupole) and a spin-1 (dipole) excitation. We design two types of quench protocols corresponding to the change of mass tensor and the application of an electric field, which are shown to excite either the individual modes or their combination. We support these findings using extensive exact diagonalization calculations of the real-time evolution of the FQH bilayer system and formulating a field-theoretic description of the quench.

{\it Model.} We consider a bilayer FQH system at total filling $\nu=2/3$ on the square torus with $N$ bosons and $N_\phi=N/\nu$ magnetic flux quanta. We label the two layers by $\sigma=\uparrow,\downarrow$, and neglect interlayer tunneling.
Hence, the number of bosons in each layer, $N^\sigma$, is conserved and we focus on the density-balanced case with pseudospin $S_z\equiv \frac{1}{2}(N^\uparrow-N^\downarrow)=0$. We assume that the bosons reside in the lowest Landau level (LLL), and their interaction is described by the Hamiltonian
\begin{eqnarray}
H = \sum_{{\bf q}}\sum_{\sigma,\sigma'=\uparrow,\downarrow} \bar{V}^{\sigma,\sigma'}_{\bf q} :{ \rho}^{\sigma}_{\bf q} { \rho}^{\sigma'}_{-{\bf q}}: \; .
\label{hamil}
\end{eqnarray} 
Here ${ \rho}^\sigma_{\bf q}=\sum_{j=1}^{N^\sigma}e^{i{\bf q}\cdot{\bf R}_j^\sigma}$ is the LLL-projected density operator in layer $\sigma$, with ${\bf R}_j^\sigma$ the $j^{\rm th}$ particle's guiding center coordinate~\cite{Prange87}, ${\bar{V}}^{\sigma,\sigma'}_{\bf q}$ is the Fourier transform of the interaction,  and $::$ denotes normal ordering. 

The Fourier transform of the interaction is a product of the Coulomb potential and the LLL form factors, ${\bar V}^{\sigma,\sigma'}_{\bf q}=V^{\sigma,\sigma'}_{\bf q}F^\sigma_{\bf q}F^{\sigma'}_{\bf q}$. The intralayer potentials are $V^{\uparrow\uparrow}_{\bf q}=V^{\downarrow\downarrow}_{\bf q}=2\pi/|{\bf q}|$, and the interlayer interaction is $V^{\uparrow\downarrow}_{\bf q}=(V^{\downarrow\uparrow}_{\bf q})^*=(2\pi/|{\bf q}|)e^{-|{\bf q}|d}e^{i{\bf q}\cdot{\bf s}}$, where $d$ is the interlayer distance, and ${\bf s}=(s_{x},s_{y})$ is the relative displacement between the bosons in different layers. Throughout this work we quote energies in units of $e^{2}/(\varepsilon\ell_{B})$, where the magnetic length $\ell_B=\sqrt{\hbar c/eB}$ and $\varepsilon$ is the dielectric constant of the host material. The quantity $d/\ell_B$ can be varied by changing the magnetic field, while ${\bf s}$ can be tuned by applying an electric field in one layer. The form factor $F^\sigma_{\bf q}=\exp[-(g_m^{\sigma})^{ab}q_a q_b \ell_B^2/4]$ depends on the band mass tensor in each layer $g_m^{\sigma}$~\cite{Haldane11} (we use Einstein's summation convention). The $2\times2$ unimodular matrix $g_m^\sigma$ measures the mass anisotropy in layer $\sigma$ which is induced, e.g., by tilting the magnetic field. In the isotropic case we have $g_m^\sigma=\id$, where $\id$ is the $2\times2$ identity matrix. 

For small interlayer distances, $d\lesssim \ell_B$, the ground state of the bosonic $\nu=2/3$ FQH bilayer is described by the Halperin $(221)$ state~\cite{Halperin83}, an incompressible fluid with total momentum ${\bf k}={\bf 0}$. At large values of $d$, the system transitions to two decoupled $\nu=1/3$ states, each being a bosonic analog of the composite fermion Fermi liquid~\cite{Halperin93,SM}. We are interested in probing the non-equilibrium behavior of the $(221)$ system using a global quench of the system's Hamiltonian. In our calculations we fix $d=0.4\ell_B$. Initially the system is in the ground state $|\Psi_0\rangle$ of $H_0\equiv H(g_m^{\uparrow,\downarrow}=\id,{\bf s}={\bf 0})$ in the $(221)$ phase. At time $t=0$, we suddenly modify the Hamiltonian $H_0\to H'$, and let the system evolve according to the Schr\"odinger equation $|\Psi(t)\rangle=e^{-iH't}|\Psi_0\rangle$. 

\begin{figure}
	\centerline{\includegraphics[width=0.98\linewidth]{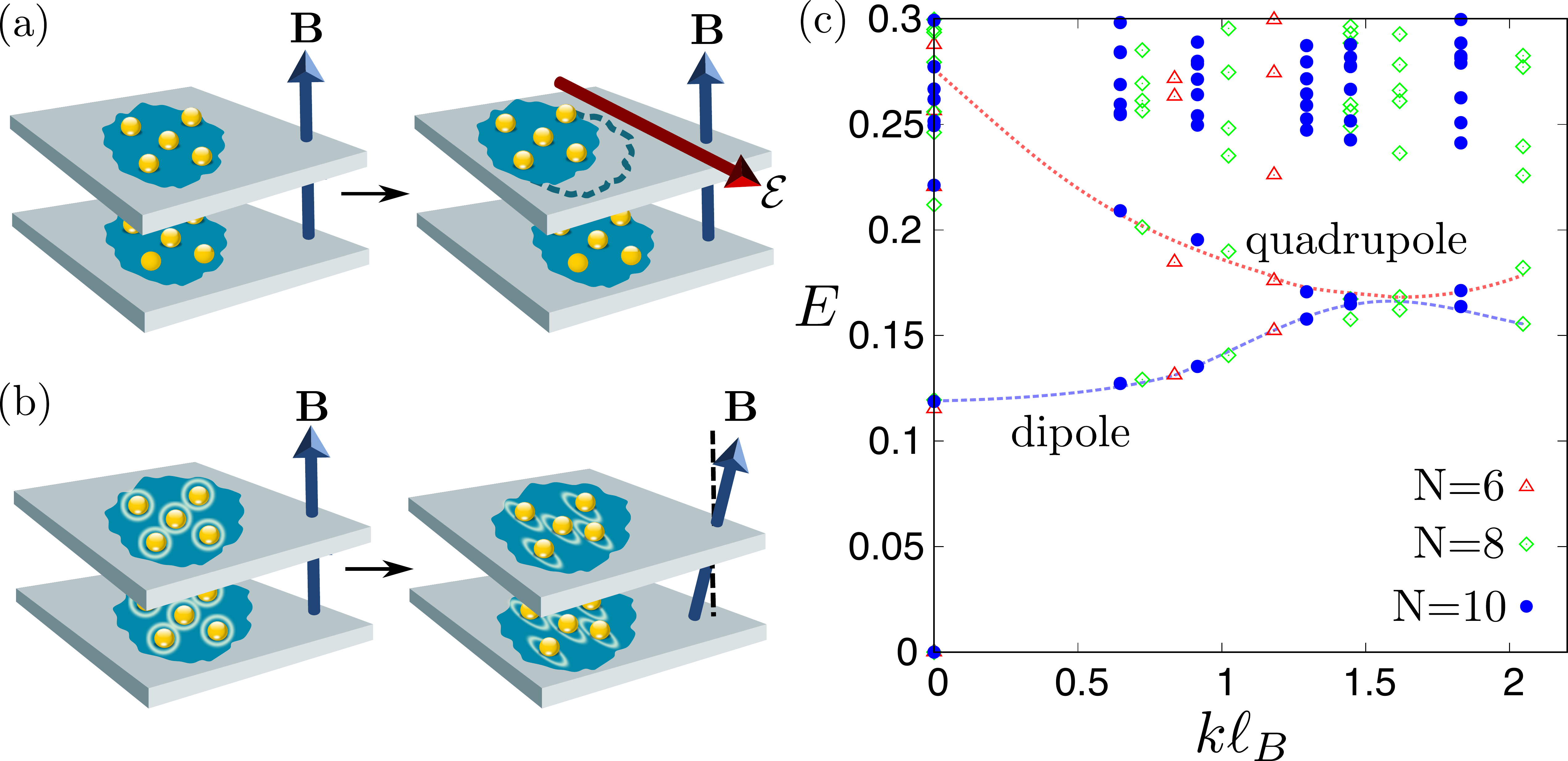}}
	\caption{(a) Instantaneous application of an electric field $\mathcal{E}$ in one layer induces dynamics in the relative displacement between FQH droplets in the two layers. (b) Instantaneous change of the mass tensors in both layers (or, equivalently, a sudden tilt of the magnetic field ${\bf B}$) induces dynamics in the intrinsic anisotropy~\cite{Liu18}, which describes the shape of flux-particle composites of the underlying FQH state. (c) Coulomb spectra of $N=6,8,10$ bosons on the torus for $d=0.4\ell_B$ showing the quadrupole and dipole collective modes. Dashed lines trace out these collective modes as a guide to the eye. }
	\label{fig1}
\end{figure}

The sudden change of the Hamiltonian defines the quench, and we consider two protocols: (i) applying electric field in a single layer [Fig.~\ref{fig1}(a)], which is equivalent to changing ${\bf s}={\bf 0}\rightarrow{\bf s}'\neq{\bf 0}$; and/or (ii) changing the mass tensor $g_m^\sigma=\id\rightarrow {g_m^{\sigma}}'\neq\id$ to add anisotropy in both layers [Fig.~\ref{fig1}(b)], where ${g_m^{\sigma}}'$ is taken to be diagonal for simplicity. We find that the essential features of post-quench dynamics are independent on precise values of ${\bf s}'$ and ${g_m^{\sigma}}'$ as long as the ground state of $H'\equiv H({g_m^{\sigma}}={g_m^{\sigma}}',{\bf s}={\bf s}')$ remains in the $(221)$ phase, which we assume below. 

The key to understanding the dynamics lies in the excited states of $H(g_m^{\sigma},{\bf s})$. A typical energy spectrum of the $(221)$ system on the torus is shown in Fig.~\ref{fig1}(c). The ground state is in the ${\bf k}={\bf 0}$ momentum sector, and there are two excitation modes above it. On the sphere, the upper mode starts from the total angular momentum $L=2$ and hence is termed a quadrupole mode~\cite{Girvin85, Girvin86}, while the lower mode starts from $L=1$ and forms a dipole excitation. We note that the dispersion of these two modes is not sensitive to the precise values of ${g_m^{\sigma}}$ and ${\bf s}$. In the language of field theory, the two modes are described using a degree of freedom that carries spin-2 and spin-1,  respectively, in the long-wavelength limit. In the context of SMA, the long-wave limits of the quadrupole and dipole modes can be obtained by acting on the ground state with ${\rho}^S_{\bf q}=({\rho}^{\uparrow}_{\bf q}+{\rho}^{\downarrow}_{\bf q})/\sqrt{2}$ and ${\rho}^{AS}_{\bf q}=({\rho}^{\uparrow}_{\bf q}-{\rho}^{\downarrow}_{\bf q})/\sqrt{2}$, respectively~\cite{Renn93,MacDonald94,Shizuya04}. As our quench protocols preserve translation symmetry, only eigenstates with ${\bf k}={\bf 0}$ are involved in the dynamics.

\begin{figure}
	\centerline{\includegraphics[width=\linewidth]{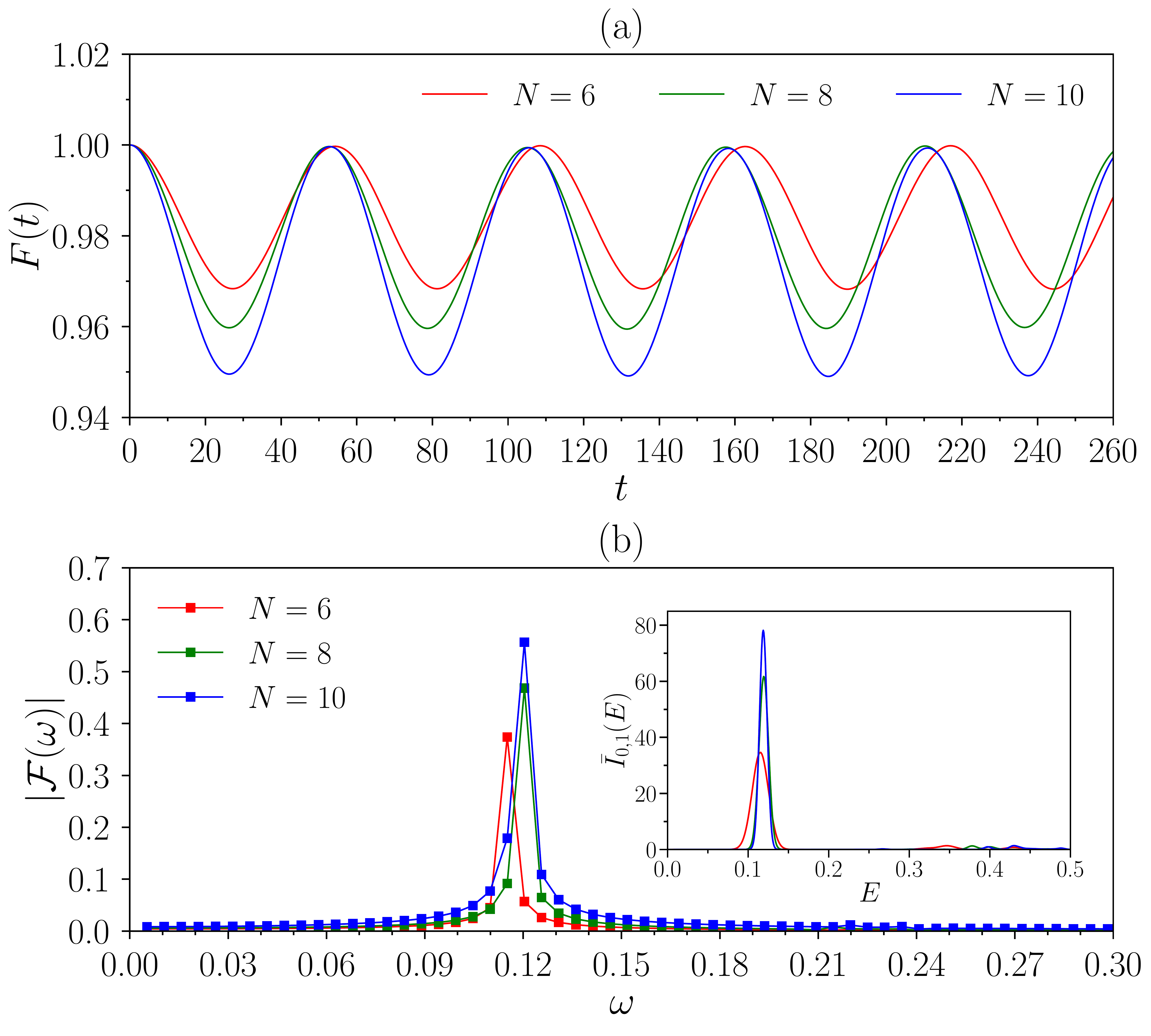}}
	\caption{(a) The fidelity $F(t)$ and (b) its discrete Fourier transform $|\mathcal{F}(\omega)|$ for the quench driven by tuning interlayer displacement from $(0,0)$ to $(s,0)$ with $s=0.1\ell_B$. The inset of (b) shows the normalized spectral function $\bar{I}_{0,1}(E)=I_{0,1}(E)/\int I_{0,1}(E) dE$ for isotropic systems with zero interlayer displacement. Markers in the main figure and curves in the inset with the same color refer to the same system size.
	}
	\label{fig2}
\end{figure}

\begin{figure*}
	\centerline{\includegraphics[width=\linewidth]{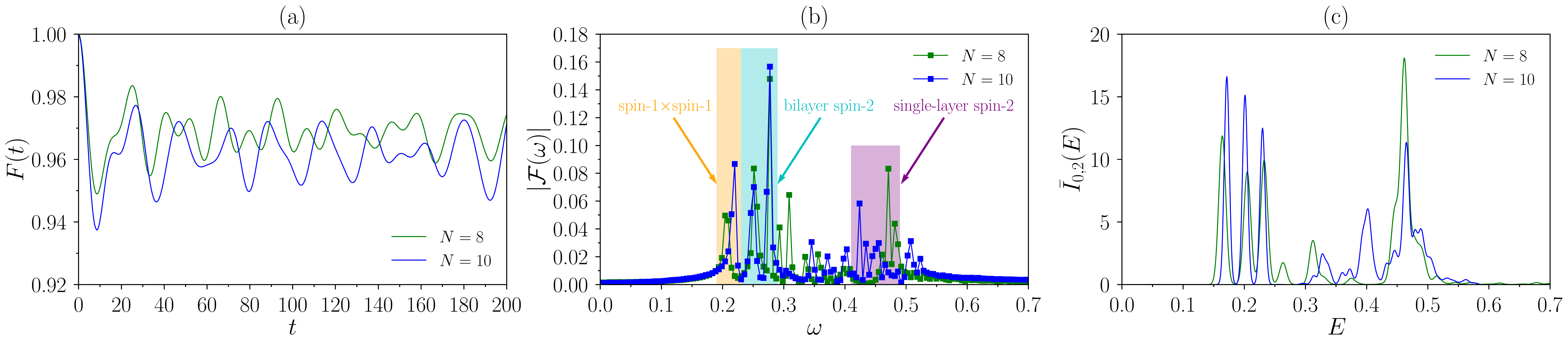}}
	\caption{(a) The fidelity $F(t)$ for the quench driven by tuning the mass tensors in both layers from $\id$ to ${\rm diag}\{\alpha,1/\alpha\}$ with $\alpha=1.3$ for $d=0.4\ell_{B}$. (b) The discrete Fourier transform $|\mathcal{F}(\omega)|$ of $F(t)$. The three types of dominant frequencies, i.e., the combination of two spin-$1$ modes (orange), the bilayer spin-$2$ graviton (ctan), and the spin-2 state in a single layer (purple), are indicated by shaded areas and arrows. (c) The normalized spectral function $\bar{I}_{0,2}(E)=I_{0,2}(E)/\int I_{0,2}(E) dE$ for isotropic systems with zero interlayer displacement (very similar data are obtained for weakly anisotropic systems with small ${\bf s}$).  
	}
	\label{fig3}
\end{figure*} 

{\it Electric-field quench.} Let us first consider the quench in which we apply an electric field instantaneously in one layer while keeping $g_m^\sigma$ in both layers isotropic. For simplicity, we consider an electric field in the $x$-direction, whose effect can be captured by changing the interlayer displacement from $(0,0)$ to $(s,0)$, with $s\neq 0$ (the electric field also lifts the degeneracy of the LLL orbitals, but this effect is negligible for the system sizes we study). We compute the post-quench fidelity $F(t)=|\langle\Psi_0|\Psi(t)\rangle|$ to monitor the dynamics. We find that $F(t)$ oscillates regularly with a single dominant frequency, as shown in Fig.~\ref{fig2}(a) for $s=0.1\ell_B$, and this frequency is almost the same for different system sizes and other small $s$. To extract this frequency, we plot the discrete Fourier transform $|\mathcal{F}(\omega)|$ of $F(t)$ in Fig.~\ref{fig2}(b). We see that $|\mathcal{F}(\omega)|$ has a sharply pronounced peak at $\omega\approx0.12$ which is in excellent agreement with the energy of the spin-$1$ dipole mode in the long-wavelength limit [see Fig.~\ref{fig1}(c)]. 

As shown in Fig.~\ref{fig1}(c), the entire dipole mode lies below the continuum of the energy spectrum. This allows us to readily identify the coherent oscillations under the applied electric field with the dipole mode. This will not be the case with other types of quenches considered below. To unambiguously identify the modes excited by a quench, we construct appropriate spectral functions and look for peaks in it. In the dipole case, we use the spectral function of an operator carrying spin-1 evaluated in the ${\bf k}={\bf 0}$ sector. A natural choice for such an operator is the $V_{0,1}$ generalized pseudopotential~\cite{Yang17a}, adapted to the bilayer case i.e.,  
$\hat{V}_{0,1}= \sum_{{\bf q}} \bar{ V}_{0,1}({\bf q}) :({\rho}^{\uparrow}_{\bf q} { \rho}^{\downarrow}_{-{\bf q}}-{\rho}^{\downarrow}_{\bf q} { \rho}^{\uparrow}_{-{\bf q}}):$ with $V_{0,1}({\bf q})\propto iq_x$. The corresponding spectral function $I_{0,1}(E)$ is 
\begin{eqnarray}
I_{0,1} (E)=\sum_j \delta(E-\epsilon_j+\epsilon_0) |\langle j | \hat V_{0,1} |0\rangle|^2.
\label{i01}
\end{eqnarray} 
Note that the bilayer $\hat{V}_{0,1}$ is defined to be antisymmetric with respect to the layer index because all layer-symmetric terms vanish for $V_{0,1}({\bf q})$. As $\hat{V}_{0,1}$ couples the ground state with excited states with $L_z=1$ (spin-$1$), the peaks in $I_{0,1}(E)$ correspond to spin-$1$ eigenstates. We show $I_{0,1}(E)$ in the inset of Fig.~\ref{fig2}(b) for isotropic systems with $s=0$ (very similar data are obtained for weakly anisotropic systems with small $s$). Indeed, $I_{0,1}(E)$ has a sharp peak at $E\approx 0.12$, agreeing with the lowest-excited state in the ${\bf k}={\bf 0}$ sector. This further confirms that the long-wave limit of the spin-$1$ dipole mode governs the electric-field-driven quench dynamics. 

{\it Mass anisotropy quench.} 
We now turn to the quench driven by mass anisotropy. In this case, we drive the quench by keeping ${\bf s}={\bf 0}$ and changing the mass tensors $g_m^{\sigma}$ in both layers from $\id$ to ${\rm diag}\{\alpha,1/\alpha\}$ with $\alpha>1$ at $t=0$. In single-layer FQH systems, the quench dynamics driven by mass anisotropy is dominated by a single spin-$2$ degree of freedom, which was identified with the long-wave limit of the GMP mode~\cite{Liu18}. Since bilayer FQH systems have multiple neutral excitations, we expect the dynamics of bilayer mass-anisotropy quench to be richer than the single-layer case.

Like in the electric-field quench, we first study the fidelity $F(t)$, shown in Fig.~\ref{fig3}(a) for $\alpha=1.3$. It is clear that $F(t)$ now oscillates with multiple frequencies. To extract the dominant frequencies we plot the discrete Fourier transform $|\mathcal{F}(\omega)|$ of $F(t)$ in Fig.~\ref{fig3}(b). Indeed we observe several pronounced peaks that are insensitive to small variations in $\alpha$. As changing the mass tensor leads to quadrupolar (spin-$2$) deformations of FQH droplets~\cite{Yang17a,Liu18}, we expect these dominant frequencies to correspond to spin-$2$ degrees of freedom in the ${\bf k}={\bf 0}$ sector of $H'$.  To substantiate this quantitatively, we utilize the spectral function of a spin-2 operator in the ${\bf k}={\bf 0}$ sector. We choose the operator $\hat{V}_{0,2} = \sum_{{\bf q}} \bar{ V}_{0,2}({\bf q}) :{ \rho}^{S}_{\bf q} {\rho}^{S}_{-{\bf q}}:$ with $V_{0,2}({\bf q})\propto q_x^2-q_y^2$, which is the bilayer generalization of the $V_{0,2}$ generalized pseudopotential~\cite{Yang17a}. Its spectral function $I_{0,2}(E)$ is defined analogously to Eq.~(\ref{i01}). As shown in Fig.~\ref{fig3}(c), the positions of peaks in $I_{0,2}$ indeed match those in $|\mathcal{F}(\omega)|$. Thus all dominant frequencies in the post-quench dynamics correspond to spin-$2$ eigenstates in the ${\bf k}={\bf 0}$ sector of $H'$. 

What is the physical interpretation of the multiple spin-$2$ states observed in the dynamics? On the one hand, the long-wave limit of the quadrupole mode, i.e., the bilayer spin-$2$ graviton, should definitely contribute. As suggested by the exact energy spectrum in Fig.~\ref{fig1}(c), the quadrupole mode approaches the energy $E\approx 0.25-0.3$ in the long-wave limit, and there are indeed corresponding sharp peaks in $|\mathcal{F}(\omega)|$ [cyan-shaded area in Fig.~\ref{fig3}(b)]. The splitting of these peaks is due to  ``fragmentation'' of the spin-$2$ graviton mode into several states in finite systems, a feature which is also observed in single-layer systems~\cite{Liu18}. On the other hand, although the long-wave limit of the spin-$1$ dipole mode cannot couple to the quench, a suitable combination of two spin-$1$ states can be excited by it. Two spin-$1$ states can form a bound state with spin-$2$, whose energy is slightly reduced from twice the spin-$1$ energy. The spectrum shown in Fig.~\ref{fig1}(c) and results of the electric-field quench suggest that the dipole mode goes to $E\approx0.12$ in the long-wave limit, thus a bound state of two dipoles, with energy $E<0.24$, could appear in the post-quench dynamics. Remarkably, we indeed observe a sharp peak at that energy in $|\mathcal{F}(\omega)|$ [orange-shaded area in Fig.~\ref{fig3}(b)].

Curiously, in addition to the bilayer spin-$2$ graviton and the bound state of two spin-$1$s, we also see peaks in $|\mathcal{F}(\omega)|$ at a much higher frequency $\omega\approx0.45-0.5$ [purple shaded area in Fig.~\ref{fig3}(b)]. In principle, higher multiples of the elementary spin-$1$ and spin-$2$ modes may be expected to appear in the dynamics, but their contribution to the spectral function is expected to be significantly reduced. Moreover, we find these higher-frequency peaks become sharper and move to higher frequencies with increasing $d$ when the two layers are progressively more decoupled~\cite{SM}. Hence, we identify this spin-$2$ excitation with that of a single-layer $\nu=1/3$ bosonic system. 

{\it Effective field theory.}  Similar to the single-layer case, the bilayer spin-$2$ graviton can be described by the bimetric theory~\cite{Gromov17}. Here we outline the effective theory describing the new collective spin-$1$ mode. The theory is a spin-$1$ counterpart of the bimetric theory with a vector degree of freedom $v_i$ that quantifies relative displacement of layers, described by the Lagrangian
\begin{equation}
\label{eq:Leff}
\mathcal L =  -\epsilon^{ij} v_i \dot{v}_j - M |v|^2 + E^-_i v_i\,,
\end{equation}
where $E_i^-$ is the difference between electric fields applied to the layers and $M$ determines the gap of the spin-$1$ mode.
The quench is simulated by suddenly switching on $E^-_i$ at $t=0$ and solving classical equations of motion \cite{franchini2015universal, Liu18}. Assuming that the quench is along $x$ direction, i.e., $E_y^-=0$, the equations of motion stemming from Eq.~\eqref{eq:Leff} are single harmonics
\begin{eqnarray}
v_x(t)=A[1-\cos(M t)], v_y(t)=A\sin(M t),
\label{delta}
\end{eqnarray} 
where the amplitude of oscillations is determined by the quench strength, $A=E_x^-/(2M)$. 

\begin{figure}[b]
	\centerline{\includegraphics[width=\linewidth]{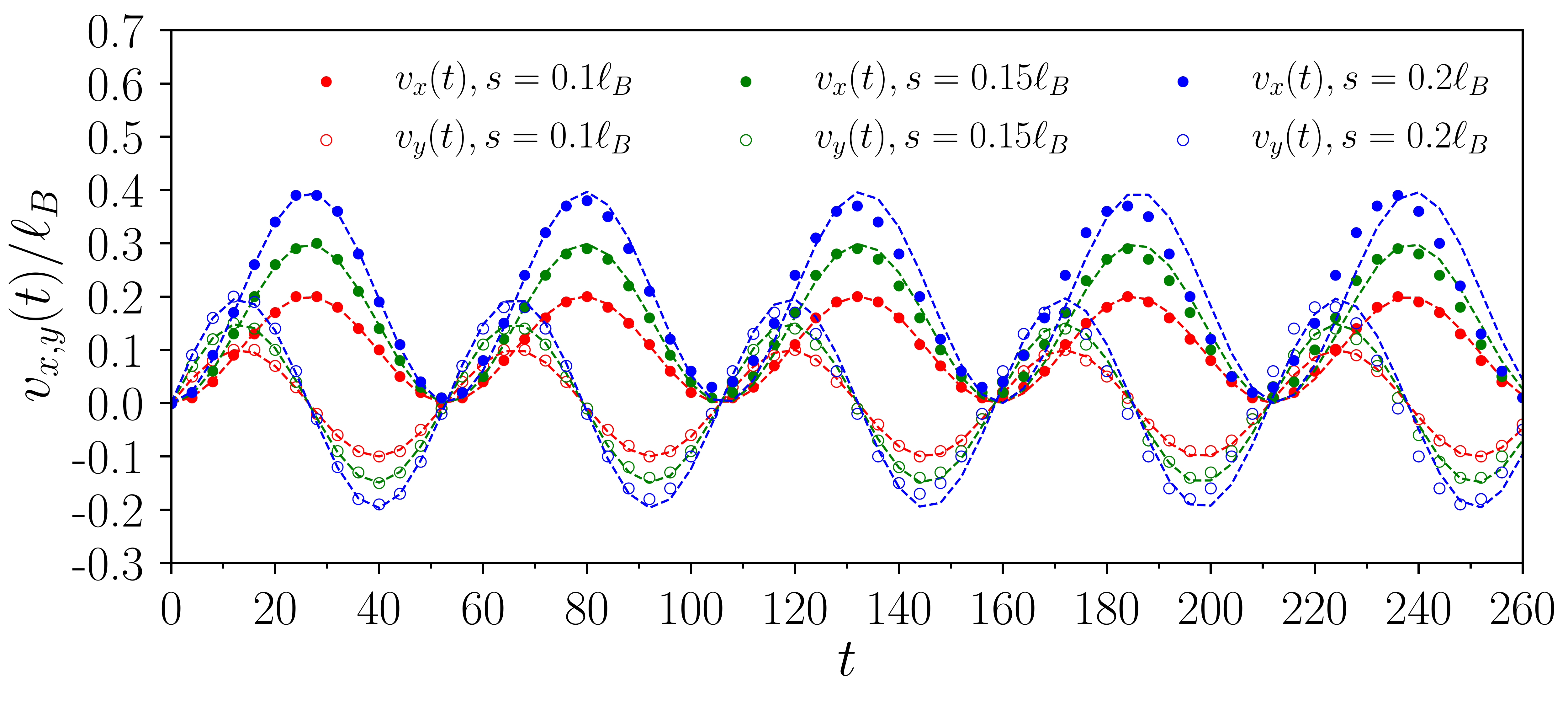}}
	\caption{Exact dynamics of the interlayer displacement ${\bm v}$ in the $x$ (dots) and $y$ directions (circles) for $N=10$ bosons after quenches driven by tuning the interlayer displacement from $(0,0)$ to $(s,0)$ with $s=0.1\ell_B,0.15\ell_B,0.2\ell_B$. The dashed curves are fits to Eq.~(\ref{delta}). 
	}
	\label{fig4}
\end{figure}

Dynamics of $v_i$ coming from Eq.~(\ref{delta}) can then compared to a numerical simulation, 
where $v_i(t)$ is determined by a brute force search over a large set of precomputed trial $(221)$ states $|\Psi_{\rm trial}({\bf s})\rangle$, i.e., the ground state of the Hamiltonian $H(g_m^{\uparrow,\downarrow}=\id,{\bf s})$. When the overlap $|\langle\Psi(t)|\Psi_{\rm trial}({\bf s})\rangle|$ is maximized (and sufficiently close to unity), we expect  $v_i(t) = s_i$. In Fig.~\ref{fig4}, we show dynamics of $v_i$ for various weak quench strengths and compared to Eq.~(\ref{delta}). 
Fitting the first oscillation in Fig.~\ref{fig4} against Eq.~(\ref{delta}), we find a remarkable agreement between numerically exact dynamics and field-theory predictions up to moderate times. The frequency $M$ returned by the fit matches the energy $\approx 0.12$ of the ${\bf k}={\bf 0}$ spin-$1$ state, and the oscillation amplitude is given by $A=2s$. With increasing quench strength or at longer times, we observe deviations from simple harmonic oscillations, which we believe is caused by effects like fragmentation of the long-wave limit of the dipole mode and the interaction between spin-$1$ states.

To describe the spin-$2$ bound state of the spin-$1$ modes we must include the interaction term, $\mathcal L_{int} \propto |v|^4$,  into Eq.~\eqref{eq:Leff}. It is then straightforward to show that $\langle v_i v_j \rangle$ behaves as a spin-$2$ mode and responds to the geometric quench, leading to an extra peak in Fig. \ref{fig3}(b).

{\it Discussion.} In this work, we explored the quench dynamics of collective modes in bilayer $\nu=2/3$ systems of bosons. We proposed and numerically simulated two quench protocols which excite neutral degrees of freedom in the system. The quench driven by an electric field applied in one layer induces oscillations of the long-wavelength limit of the spin-$1$ dipole collective mode. More interestingly, the quench driven by mass anisotropy not only activates the spin-$2$ quadrupole mode but also single-layer spin-$2$ excitation and a combination of two spin-$1$ dipole modes. While in this paper we presented results for systems of bosons, all of our conclusions also hold for FQH systems of fermions.

Direct access to the spin-$1$ mode in the spectrum of bilayer states suggests a variety of new problems related to the geometric aspects of FQH and exact calculations of correlation functions \cite{Nguyen17, Gromov15, Gromov14a, Gromov14,douglas2010bergman,ferrari2014fqhe,Can14,Haldane11}. Namely, which correlations functions are sensitive to the mode, and can any of them be computed with the help of existing methods?

Our quench protocols provide an opportunity to experimentally measure the collective modes of FQH states at long wavelengths in a way that complements the inelastic light scattering~\cite{Pinczuk93, Platzman96, Kang01, Kukushkin09} and current noise measurements~\cite{Joglekar04}. In fact, the quench protocols proposed in this work, in particular the counterflow electric field, can be implemented with the existing technology. The main challenge would be resolving the dynamics on short time scales in solid-state materials, which would be naturally resolved in other platforms, e.g., cold atoms in optical lattices forming a fractional Chern insulator~\cite{parameswaran2013review,liu2013review, Titusreview}. Our results are also of direct relevance to more complex FQH systems with non-Abelian topological order, which also host multiple types of neutral excitations~\cite{Moller11, Bonderson11}. It would be interesting to design quench protocols and effective theories to probe different collective modes at long wavelengths as well as the combination (or interaction) between them.  
 
{\sl Acknowledgments.} Z.~L. is supported by the National Natural Science Foundation of China through Grant No.~11974014. Some of the numerical calculations reported in this work were carried out on the Nandadevi supercomputer, which is maintained and supported by the Institute of Mathematical Science’s High Performance Computing center. Z.P. acknowledges support by the Leverhulme Trust Research Leadership Award RL-2019-015. A.G. is supported by the Brown University.

\bibliography{biblio_fqhe}

\onecolumngrid 
\newpage

\section{Supplementary Material}
\setcounter{subsection}{0}
\setcounter{equation}{0}
\setcounter{figure}{0}
\renewcommand{\theequation}{S\arabic{equation}}
\renewcommand{\thefigure}{S\arabic{figure}}
\renewcommand{\thesubsection}{S\arabic{subsection}}

In this Supplementary Material, we provide additional numerical data for the quench dynamics of $\nu=2/3$ bilayer bosons. We then use the composite-fermion (CF) exciton and single-mode approximation (SMA) to predict the two neutral collective modes of the system -- the quadrupole mode and the dipole mode. By comparing with the spectra obtained from exact diagonalization, we find that the CF exciton describes the collective modes much better than SMA at finite momenta. We present numerical evidence that the bilayer forms a compressible state at large layer separations, consisting of two decoupled $\nu=1/3$ CF liquids of bosons. Finally, we outline an effective theory to describe the bound state of two spin-$1$ degrees of freedom.

\section{Mass-anisotropy quench for various interlayer distances}
Here we present additional numerical data for mass-anisotropy quenches in Coulomb interacting systems at various interlayer distances $d$. As in the main text, the initial state is chosen to be the isotropic Coulomb ground state with zero interlayer displacement ${\bf s}$. The mass tensors in both layers are suddenly tuned from $\id$ to ${\rm diag}\{\alpha,1/\alpha\}$ to drive the quench, while the interlayer displacement is kept at zero.

\begin{figure*}
	\centerline{\includegraphics[width=\linewidth]{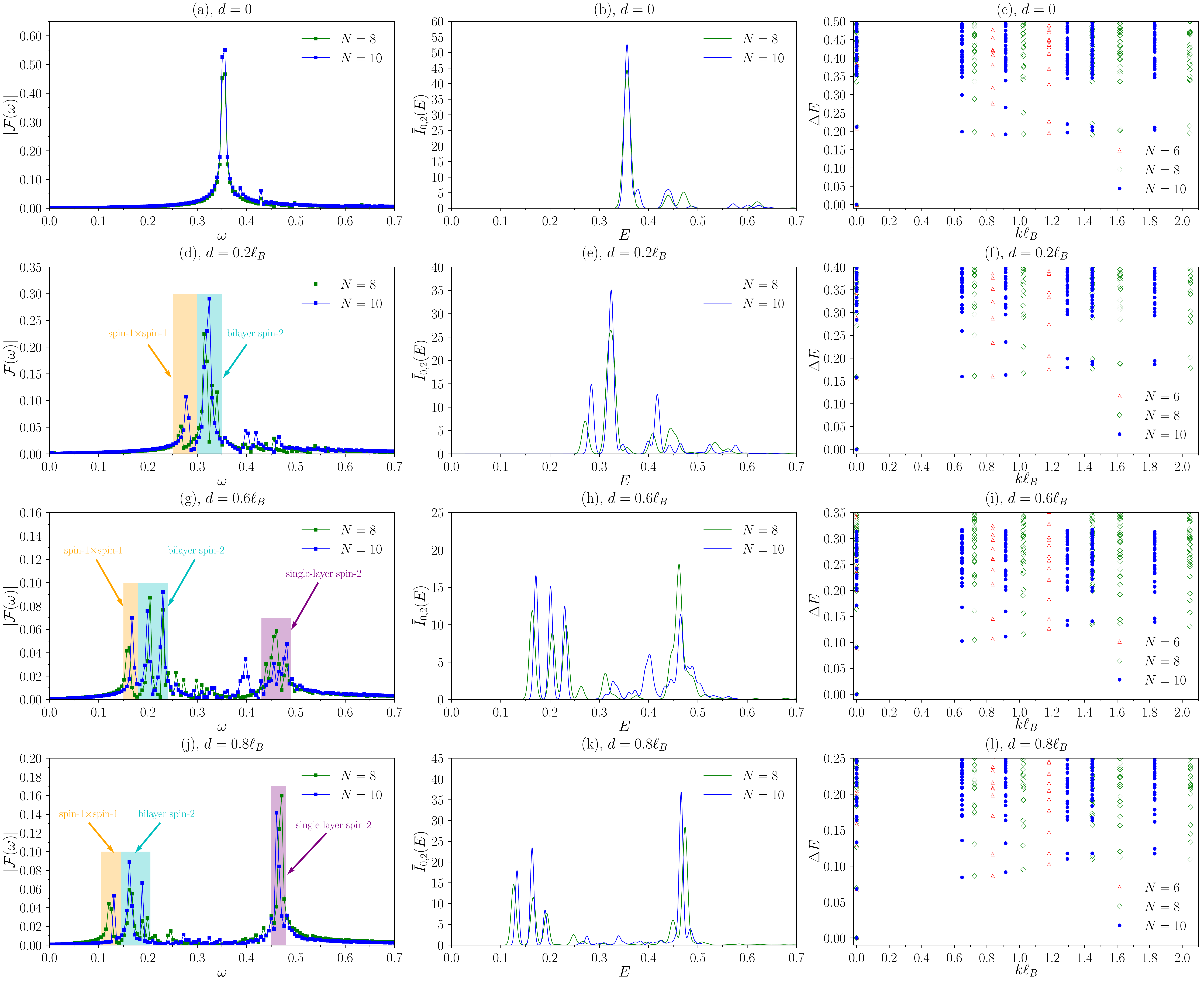}}
	\caption{(a,d,g,j) The discrete Fourier transform $|\mathcal{F}(\omega)|$ of the fidelity for the quench driven by tuning the mass tensors in both layers from $\id$ to ${\rm diag}\{\alpha,1/\alpha\}$ with $\alpha=1.3$. (b,e,h,k) The normalized spectral function $\bar{I}_{0,2}(E)=I_{0,2}(E)/\int I_{0,2}(E) dE$ for isotropic systems with zero interlayer displacement. (c,f,i,l) The exact energy spectrum. Here we consider $N=6,8,10$ Coulomb interacting bosons at (a-c) $d=0$, (d-f) $d=0.2\ell_B$, (g-i) $d=0.6\ell_B$ and (j-l) $d=0.8\ell_B$. }
	\label{smfig1}
\end{figure*}

In Fig.~\ref{smfig1}, we show the Fourier transform of post-quench fidelity, the spectral function $I_{0,2}$, and the exact Coulomb spectrum at $d/\ell_B=0,0.2,0.6$ and $0.8$. The energy spectrum has two common features for these different values of $d$: (i) the ground state is in the ${\bf k}={\bf 0}$ momentum sector and (ii) there are clear quadrupole and dipole collective modes. The spectra thus suggest that the system is in the $(221)$ phase for $d\leq \ell_{B}$. For all the values of $d$ considered here, we find clear evidence that the quench induces coherent dynamics of not only the bilayer spin-$2$ graviton (the cyan-shaded areas in Fig.~\ref{smfig1}), but also the bound state of two spin-$1$s, denoted spin-$1\times$spin-$1$, which carries spin-$2$ (the orange-shaded areas in Fig.~\ref{smfig1}). We arrive at this conclusion by comparing the peak positions in the Fourier transform $|\mathcal{F}(\omega)|$ of the post-quench fidelity and the $k\rightarrow0$ limit of the collective modes in the exact spectrum. In contrast, at $d=0$, we only observe a single sharp peak in $|\mathcal{F}(\omega)|$, which matches the energy of the bilayer spin-$2$ graviton [Fig.~\ref{smfig1}(a)]. However, because the energies of the bilayer spin-$2$ graviton and the spin-$1\times$spin-$1$ bound state should be close to each other in this case (the graviton energy is $E\approx0.35-0.4$ and the spin-$1$ energy is $E\approx0.21$) [Fig.~\ref{smfig1}(c)], this peak presumably also includes a contribution from the spin-$1\times$spin-$1$ bound state.

\begin{figure*}
	\centerline{\includegraphics[width=\linewidth]{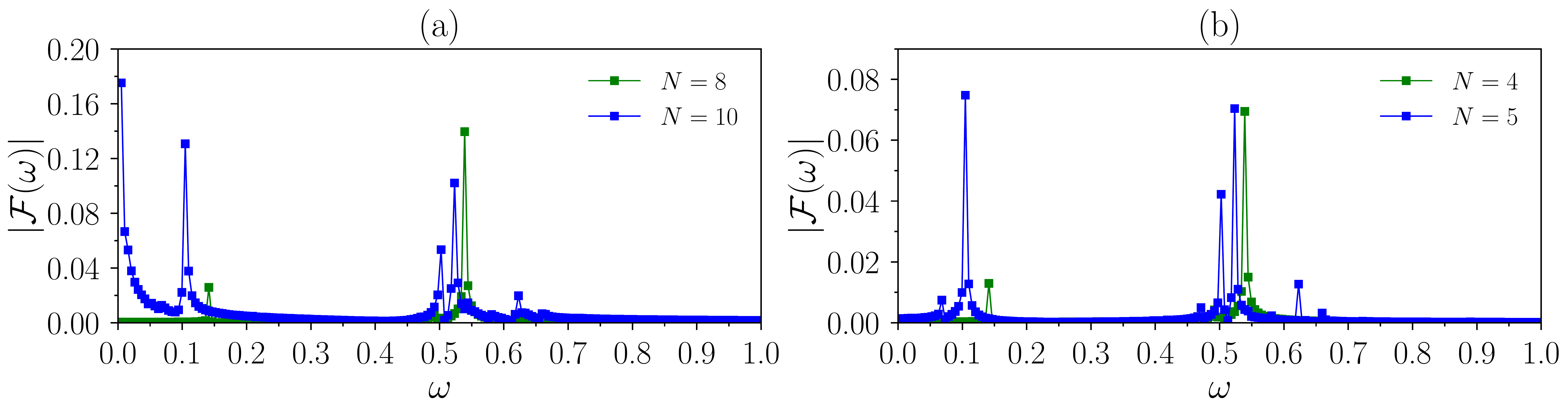}}
	\caption{(a) The discrete Fourier transform $|\mathcal{F}(\omega)|$ of the fidelity for the quench driven by tuning the mass tensors in both layers from $\id$ to ${\rm diag}\{\alpha,1/\alpha\}$ with $\alpha=1.3$. Here we consider $N=8,10$ bilayer $\nu=2/3$ Coulomb interacting bosons at $d=5\ell_B$. (b) The discrete Fourier transform $|\mathcal{F}(\omega)|$ of the fidelity for the quench driven by tuning the mass tensor from $\id$ to ${\rm diag}\{\alpha,1/\alpha\}$ with $\alpha=1.3$. Here we consider $N=4,5$ Coulomb interacting bosons for a single-layer $\nu=1/3$ system. }
	\label{smfig2}
\end{figure*} 

With increasing $d$, we find that a third dominant frequency in $|\mathcal{F}(\omega)|$ emerges [the purple-shaded areas in Fig.~\ref{smfig1}]. As shown in the main text, peaks appear at $E\approx0.45$ at $d=0.4\ell_B$, with comparable heights to those corresponding to the bilayer spin-$2$ graviton and the spin-$1\times$spin-$1$ bound state. With increasing $d$, these peaks become sharper and slowly move towards a higher frequency [see Figs.~\ref{smfig1}(g) and \ref{smfig1}(j)]. We find that these peaks exist even at very large $d$ where the system is no longer in the $(221)$ phase, suggesting that they are unrelated to excitations of the $(221)$ state. To understand this high-energy degree of freedom, which appears only at relatively large $d$, we track its position in $|\mathcal{F}(\omega)|$ up to $d=5\ell_B$. In this case, the corresponding peaks have moved to $E\approx0.5-0.55$ [Fig.~\ref{smfig2}(a)]. For such a large $d$ the two layers are nearly decoupled and each layer forms a $\nu=1/3$ bosonic state. Thus, for comparison, we study the mass-anisotropy quench in single-layer $\nu=1/3$ bosonic systems. Similar to the bilayer case, we change the mass tensor of the single-layer system from $\id$ to ${\rm diag}\{\alpha,1/\alpha\}$ with $\alpha=1.3$ to drive the quench, with the initial state chosen as the single-layer isotropic Coulomb ground state. Note that the initial state is chosen as the global ground state, i.e., it may not necessarily have ${\bf k}={\bf 0}$. Remarkably, we also observe pronounced peaks at $E\approx0.5-0.55$ in $|\mathcal{F}(\omega)|$ of quenches in single-layer systems [Fig.~\ref{smfig2}(b)]. This confirms that the high-energy degree of freedom, which starts to emerge around $d\approx 0.5\ell_B$ and exists all the way to the decoupled limit when $d\rightarrow\infty$, is a spin-$2$ excitation within a single layer. As we show below,  the bilayer system in the decoupled limit is two copies of the $\nu=1/3$ CF liquid (CFL) of bosons, so this spin-$2$ excitation is expected to be a geometric distortion of the bosonic CFL. Remarkably, signatures of this single-layer excitation could be detected in the system's dynamical response even across the quantum phase transition driven by changing $d$.

We also notice that some dominant frequencies of the bilayer system in the large-$d$ limit are absent in the dynamics of the corresponding single-layer system (Fig.~\ref{smfig2}). These frequencies correspond to bilayer eigenstates that are combinations of two single-layer states with different momenta from that of the single-layer ground state, which we cannot probe in the single-layer quench due to the momentum-preserving feature of our quench protocol.

We have also checked the electric-field quench of the $(221)$ system for various $d$ including $d=0$. The results are very similar to those shown in the main text for $d=0.4\ell_B$, i.e., the dynamics is governed by a single frequency corresponding to the spin-$1$ long-wave limit of the dipole mode.

\section{Composite-fermion exciton versus single-mode approximation}

In this section, we discuss two approximate constructions of the collective modes for bilayer bosonic systems at a total filling of $\nu=2/3$. We shall describe the construction of the wave function of the modes using both the composite fermion theory~\cite{Jain89} and the single-mode approximation (SMA)~\cite{Girvin85, Girvin86}. The CF theory gives a good description of the modes at all values of the dimensionless wave vector $k\ell_{B}$ for small layer separations $d$~\cite{Dev92, Jain97, Jain07, Balram13}. On the other hand, the SMA provides a good description of the collective modes for small $d$ only in the long-wavelength limit, i.e., $k\ell_{B}\rightarrow 0$. The CF states are most readily constructed in Haldane's spherical geometry~\cite{Haldane83}. Unless otherwise stated, all our calculations in this section are carried out in spherical geometry. For the sake of completeness, we will provide a primer on the CF theory next.

\subsection{Primer on the composite fermion theory}
\label{subsec: CF_primer}
A vast majority of the FQHE phenomenology in the lowest Landau level (LLL) is captured in terms of emergent topological particles called composite fermions, which are bound states of electrons and an even number ($2p$) of quantized vortices~\cite{Jain89, Jain07}. The CF theory postulates that a system of interacting electrons at filling factor $\nu=\nu^{*}/(2p\nu^{*}+1)$ of the LLL can be mapped onto a system of weakly interacting composite fermions at filling factor $\nu^{*}$ of CF-LLs (termed $\Lambda$Ls). In particular, integer filling of CF-LLs, i.e $\nu^{*}=n$, leads to FQHE of electrons at $\nu=n/(2n+1)$. The mapping to integer quantum Hall effect (IQHE) leads to the following Jain wave functions for FQHE states~\cite{Jain89}:
\begin{equation}
\Psi^{\alpha}_{\nu=\frac{\nu^{*}}{2\nu^{*}+ 1}} =
\mathcal{P}_{\rm LLL} \Phi_{1}^{2}\Phi^{\alpha}_{\nu^{*}}. 
\label{eq_Jain_CF_wf}
\end{equation}
Here $\alpha$ labels the different eigenstates (to keep the notation simple, we shall suppress the label $\alpha$ from here on in), $\Phi_{\nu^{*}}$ is the wave function of non-interacting electrons at $\nu^{*}$ and $\mathcal{P}_{\rm LLL}$ implements projection to the LLL. Throughout this section, we carry out projection to the LLL using the Jain-Kamilla method~\cite{Jain97, Jain97b}, details of which can be found in the literature~\cite{Moller05, Jain07, Davenport12, Balram15a}. This projection scheme, which is directly applicable to only fermionic states, allows us to access fairly large system sizes well beyond the reach of exact diagonalization.

The Jain wave functions of Eq.~(\ref{eq_Jain_CF_wf}) can be readily generalized to multi-component systems, where the components could refer to spin, orbital, layer, valley, subband, etc. degrees of freedom. Let us first consider the case of spinful systems. We first write the net filling factor of CFs as $\nu^{*}=\nu_{\uparrow}^{*}+\nu_{\downarrow}^{*}$, where $\uparrow$ and $\downarrow$ denote the up and down spins respectively. The Slater determinant $\Phi_{\nu^{*}} \equiv \Phi_{\nu_{\uparrow}^{*},\nu_{\downarrow}^{*}}$ is then simply given as a product of two Slater determinants one for each spin, i.e., $\Phi_{\nu^{*}}=\Phi_{\nu_{\uparrow}^{*}}\Phi_{\nu_{\downarrow}^{*}}$. This leads us to the following Jain wave functions for spinful electrons in the LLL:
\begin{equation}
\Psi_{\nu=\frac{(\nu_{\uparrow}^{*}+\nu_{\downarrow}^{*})}{2(\nu_{\uparrow}^{*}+\nu_{\downarrow}^{*})+ 1}} =
\mathcal{P}_{\rm LLL} \Phi_{1}^{2}(\{z\})\Phi_{\nu_{\uparrow}^{*}}(\{z^{\uparrow}\})\Phi_{\nu_{\downarrow}^{*}}(\{z^{\downarrow}\}). 
\label{eq_Jain_CF_wf_spin}
\end{equation}
At a given filling factor, one can construct states with different spin-polarizations. The spin-phase diagram of many fractional quantum Hall states in the LLL has been worked out in detail using the above wave functions~\cite{Wu93, Park98, Jain07, Liu14, Balram15, Balram15a, Balram15c, Balram17}. The above wave functions are also applicable to bilayer systems in the limit where the layer separation $d/\ell_{B} \rightarrow 0$.

For bilayer systems (with the top and bottom layers denoted by $\uparrow$ and $\downarrow$) with a finite layer separation we consider the following class of CF wave functions~\cite{Scarola01b}:
\begin{equation}
\Psi_{\nu=\frac{2\nu'}{m\nu'+ 1}} =
\prod_{i,j}(z_{i}^{\uparrow}-z_{j}^{\downarrow})^{m} \Phi_{\nu'}(\{z^{\uparrow}\})\Phi_{\nu'}(\{z^{\downarrow}\}), 
\label{eq_Jain_CF_wf_bilayer}
\end{equation}
where we have assumed that both layers are at the same filling $\nu'$ and $m$ is the number of interlayer zeroes. The Halperin $(l,l,m)$ states~\cite{Halperin83} are obtained as a special case of Eq.~(\ref{eq_Jain_CF_wf_bilayer}) when each of the layers is at a Laughlin filling of $\nu'=1/l$~\cite{Laughlin83}. A detailed phase diagram of composite fermion states in bilayer systems was recently worked out in Ref.~\cite{Faugno20}. 

Analogous states for bosons can be constructed by dividing the above fermionic wave function by the full Jastrow factor $\Phi_{1}(\{z\}) = \prod_{i<j}(z_{i}-z_{j})$. Strictly speaking, the bosonic states $\mathcal{P}_{\rm LLL}\Phi_{1}\Phi_{n}$ and $\Phi^{-1}_{1}[\mathcal{P}_{\rm LLL}\Phi^{2}_{1}\Phi_{n}]$ differ in the details of how the projection to the LLL is implemented. We expect that such details do not significantly change the nature of the state~\cite{Balram16b}. Moreover, it is only for the state $\Phi^{-1}_{1}[\mathcal{P}_{\rm LLL}\Phi^{2}_{1}\Phi_{n}]$ that the Jain-Kamilla projection method can be applied, which then allows us to access large system sizes as we shall show below. 

\begin{figure}[t]
\begin{center}
\includegraphics[width=0.47\textwidth,height=0.31\textwidth]{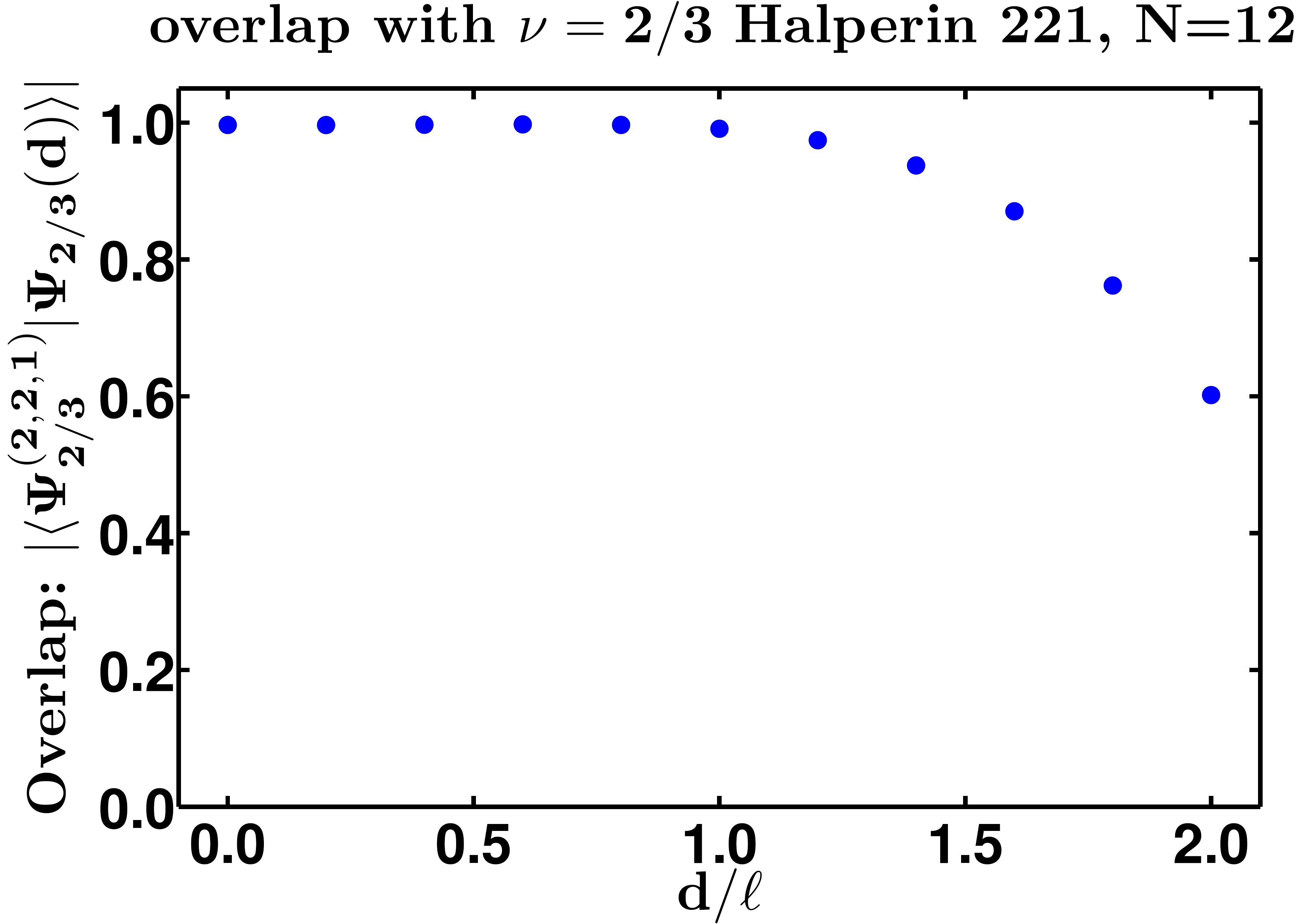} 
\caption{Overlap of the exact lowest Landau level Coulomb ground state with the Halperin $(221)$ state for bosons at $\nu=2/3$ for $N=12$ bosons at flux $2Q=16$ as a function of the layer separation $d/\ell_{B}$. }
\label{smfig6}
\end{center}
\end{figure}

Of particular interest to us is the state of bosons at total filling $\nu=2/3$. Using the above prescription, the wave function of this system is given by
\begin{equation}
\Psi_{2/3} = \Phi_{1}\Phi_{1\uparrow,1\downarrow} = \prod_{i<j}(z_{i}-z_{j})\prod_{i<j}(z^{\uparrow}_{i}-z^{\uparrow}_{j})\prod_{i<j}(z^{\downarrow}_{i}-z^{\downarrow}_{j}) = \prod_{i<j}(z^{\uparrow}_{i}-z^{\uparrow}_{j})^{2}\prod_{i<j}(z^{\downarrow}_{i}-z^{\downarrow}_{j})^{2} \prod_{i,j}(z^{\uparrow}_{i}-z^{\downarrow}_{j}) \equiv \Psi_{2/3}^{\rm Halperin221}.
 \label{eq_Jain_2_3_ss_Halperin_221}
\end{equation}
In Fig.~\ref{smfig6} we show the overlap of the exact lowest Landau level Coulomb ground state with the Halperin $(221)$ state for $N=12$ bosons as a function of the layer separation $d/\ell_{B}$. The Halperin $(221)$ state gives a good description of the ground state at small to intermediate layer separations $d$. Its overlap with the exact Coulomb ground state monotonically decreases with $d$. Although the exact lowest Landau level Coulomb ground state at finite $d$ does not have a good total pseudospin $S$, the Halperin $(221)$ is a pseudospin-singlet, i.e., has $S=0$. For small values of the anisotropy and electric fields, i.e, weak quenches, we expect the state of the system to be in the same phase as described by the above wave function~\cite{Balram16}.

\subsection{Collective modes from the composite fermion theory}
\label{subsec: collective_modes_CF_221}
\begin{figure}[t]
\begin{center}
\includegraphics[width=0.97\textwidth,height=0.11\textwidth]{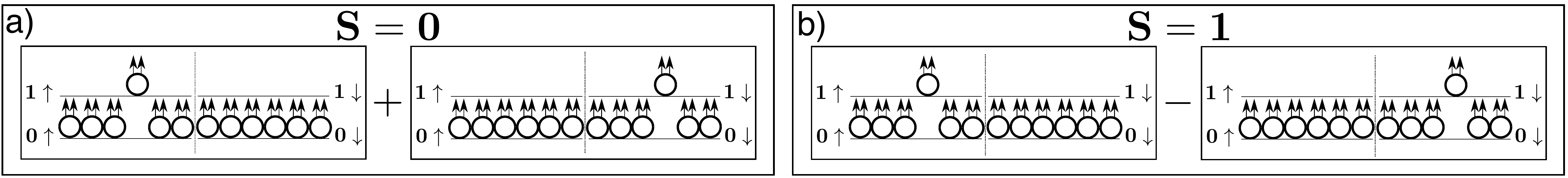} 
\caption{Schematic representation of the symmetric [left panel a)] and anti-symmetric [right panel b)] collective modes arising from CF excitons in the bilayer system of bosons (fermions) at total filling of $\nu=2/3$ ($\nu=2/5$) .}
\label{smfig7}
\end{center}
\end{figure}

For the bosonic bilayer system at $\nu=2/3$ whose ground state lies in the phase described by the wave function of Eq.~(\ref{eq_Jain_2_3_ss_Halperin_221}), the low-energy neutral excitations are obtained by creating a single CF exciton (a particle-hole pair of composite fermions). We expect to see two gapped collective modes that arise from a symmetric and anti-symmetric combination of CF exciton states in the two layers (see Fig.~\ref{smfig7}). The symmetric mode, which carries total pseudospin $S=0$, starts from total orbital angular momentum $L=2$ in the spherical geometry and hence is termed a ``quadrupole" mode (The symmetric state at $L=1$ is killed upon projection to the LLL.). The anti-symmetric mode, which carries total pseudospin $S=1$, starts from $L=1$ in the spherical geometry and hence is termed a ``dipole" mode.

The CF theory allows us to write down wave functions for these modes which enables a calculation of their energies for large system sizes. The schematic shown in Fig.~\ref{smfig7}(a) corresponds to the wave function:
\begin{equation}
\Psi^{{\rm S=0,L}}_{2/3} = \left[ \Phi_{1} \left( \left\{ z \right\} \right) \right]^{-1} \left[\mathcal{P}_{\rm LLL} \Phi^{2}_{1}(\{z\})\left(\Phi^{{\rm exciton,L}}_{1,\uparrow}(\{z^{\uparrow}\})\Phi_{1,\downarrow}(\{z^{\downarrow}\}) + \Phi_{1,\uparrow}(\{z^{\uparrow}\})\Phi^{{\rm exciton,L}}_{1,\downarrow}(\{z^{\downarrow}\}) \right) \right],
\label{eq:2/3_ss_symm_exciton}
\end{equation}
and that in Fig.~\ref{smfig7}(b) corresponds to
\begin{equation}
\Psi^{{\rm S=1,L}}_{2/3} = \left[ \Phi_{1} \left( \left\{ z \right\} \right) \right]^{-1} \left[\mathcal{P}_{\rm LLL} \Phi^{2}_{1}(\{z\})\left(\Phi^{{\rm exciton,L}}_{1,\uparrow}(\{z^{\uparrow}\})\Phi_{1,\downarrow}(\{z^{\downarrow}\}) - \Phi_{1,\uparrow}(\{z^{\uparrow}\})\Phi^{{\rm exciton,L}}_{1,\downarrow}(\{z^{\downarrow}\}) \right) \right].
\label{eq:2/3_ss_antisymm_exciton}
\end{equation}
The state $\Phi^{{\rm exciton,L}}_{1}$ denotes the wave function of a particle-hole pair at $\nu=1$ with total orbital angular momentum $L$. 

Two particles in the same layer interact via the intralayer Coulomb interaction $1/r$ while two particles in different layers interact via the interlayer Coulomb interaction $1/\sqrt{r^2+d^2}$. For $d=0$, the total pseudospin $S$ is a good quantum number and can be used to label the states. For $d>0$, the total pseudospin $S$ is not a good quantum number. However, we could still use the above states as variational states to capture the dispersion. We have evaluated the energies of the above modes using the Metropolis Monte Carlo method for various systems with different layer separations. 

\begin{figure*}[t]
\begin{center}
\includegraphics[width=0.47\textwidth,height=0.31\textwidth]{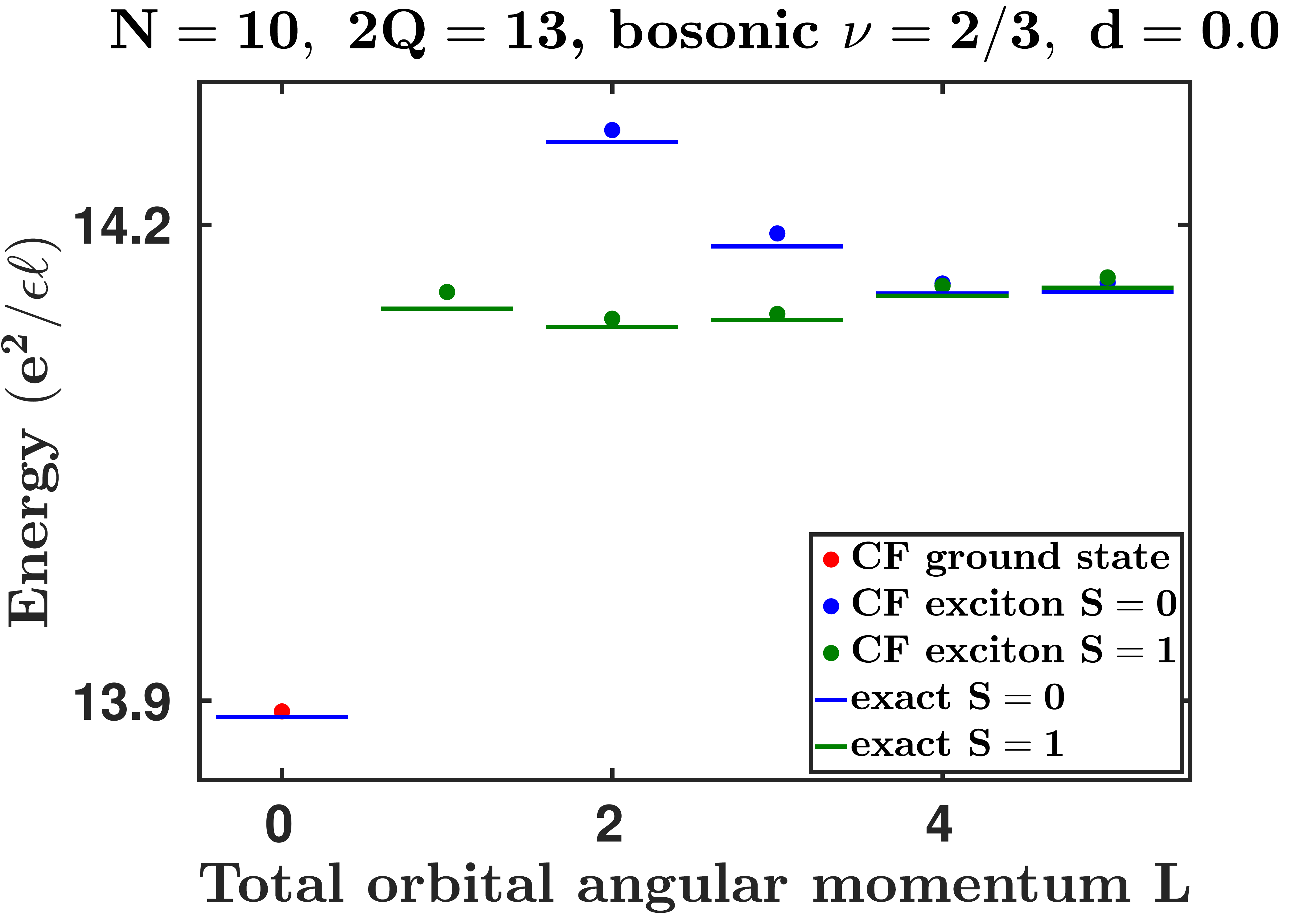} 
\includegraphics[width=0.47\textwidth,height=0.31\textwidth]{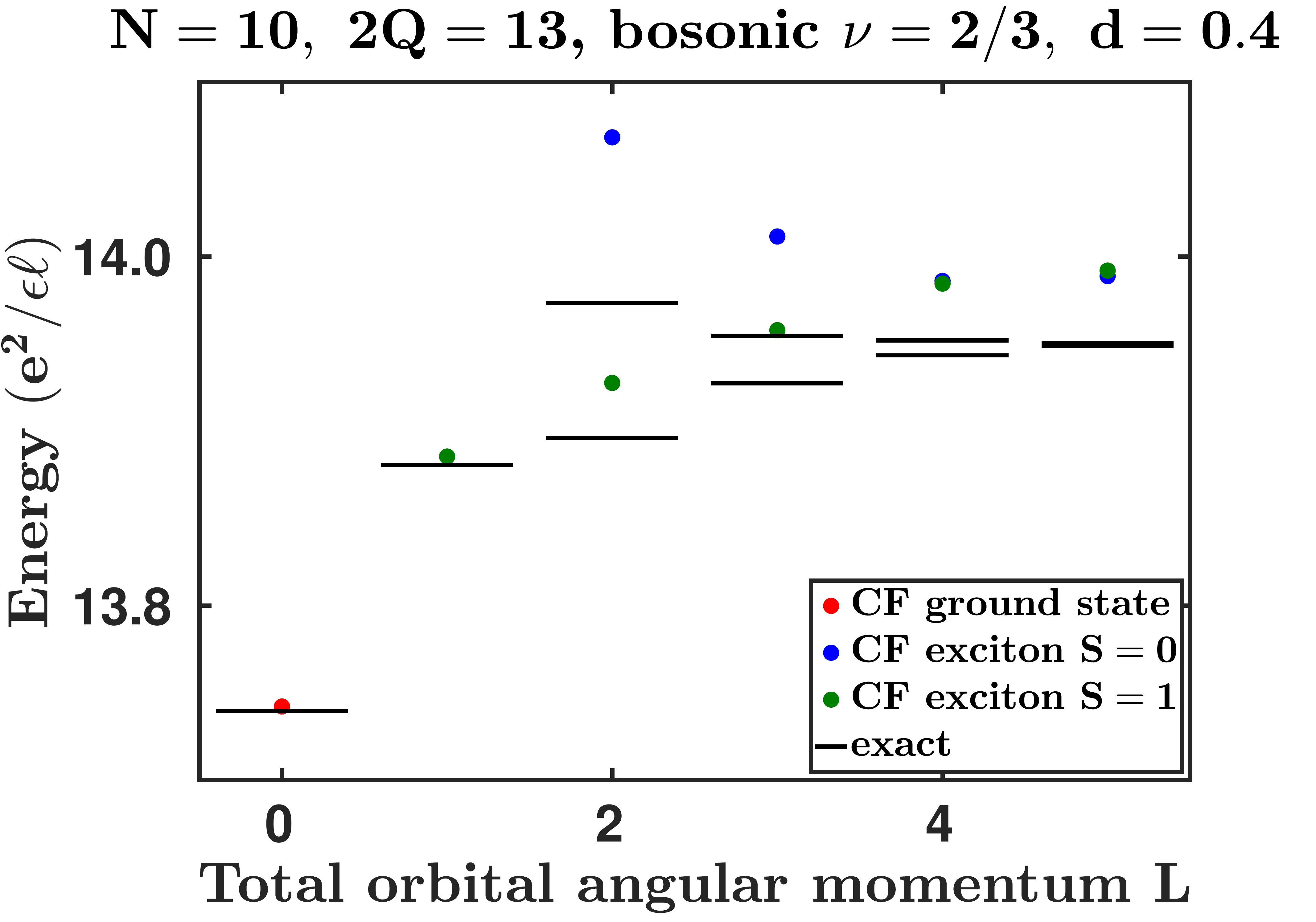} 
\caption{Exact (dashes) and composite fermion (dots) spectra for $N=10$ bosons on the sphere at a layer separation of $d/\ell_{B}=0$ (left panel) and $0.4$ (right panel). The bosonic $2/3$ Halperin $(221)$ state shown in red has total orbital angular momentum $L=0$ and total pseudospin angular momentum $S=0$. For the exact results, we show the (i) lowest energy states at $L=0=S$ and $L=1=S$ and the lowest energy states with $S=0$ and $S=1$ at $L=2,3,\cdots,N/2$ for $d=0$ and (ii) lowest energy states at $L=0,1$ and the two lowest energy states at $L=2,3,\cdots,N/2$ for $d>0$.}
\label{smfig8}
\end{center}
\end{figure*}

We first compare the exact energies (obtained from brute-force diagonalization) of a small system of $N=10$ bosons against the above variational states in Fig.~\ref{smfig8}. At $d=0$, we find that the CF modes give an excellent description of the low-energy excitations seen in the exact Coulomb spectra. However, with increasing layer-separation, the mismatch between the exact energies and those ascertained from the CF modes increases. Thus, the two CF modes are expected to have good variational energies only in the regime of small layer separations, i.e., $d/\ell_{B} < 1$.

\begin{figure*}[t]
\begin{center}
\includegraphics[width=0.47\textwidth,height=0.31\textwidth]{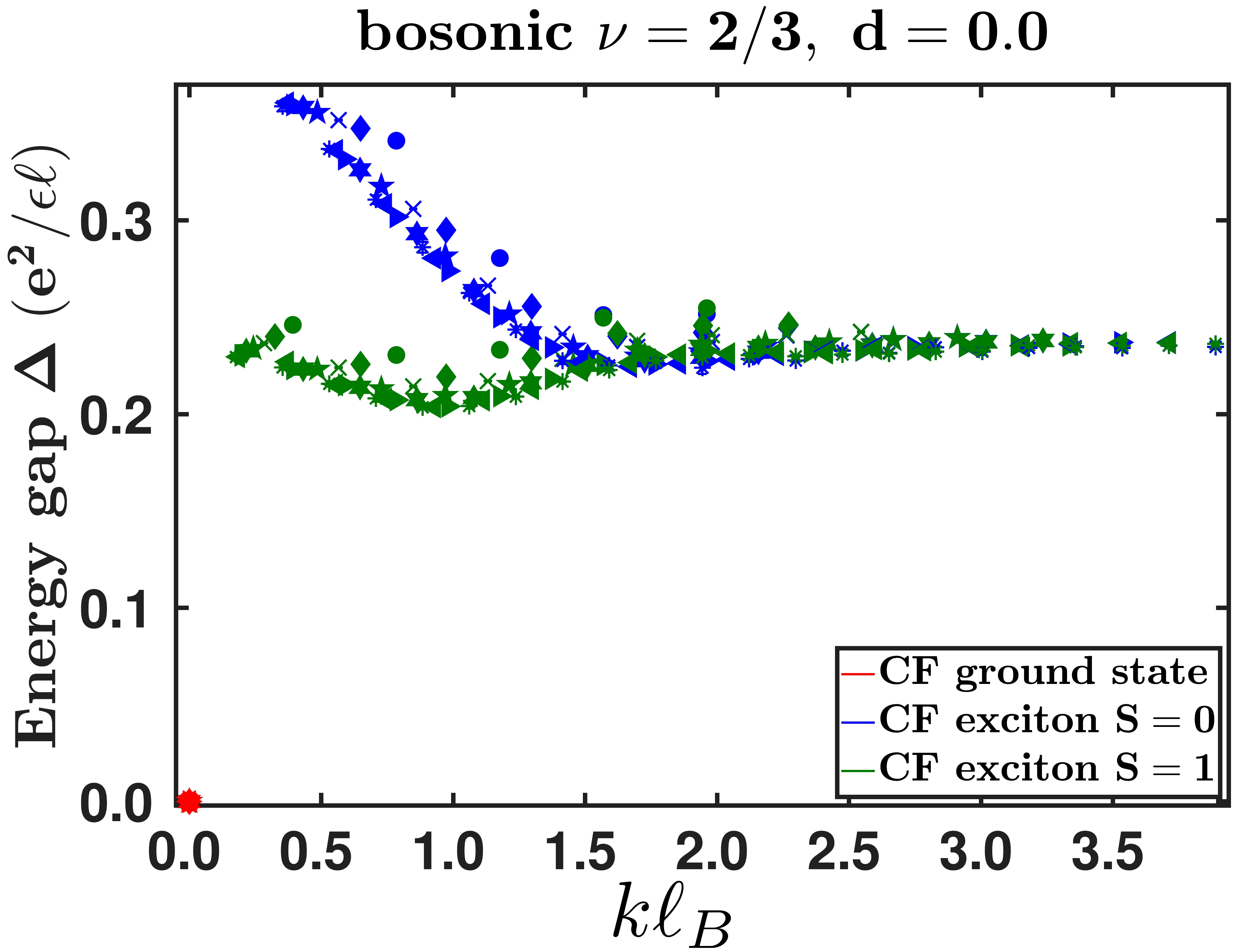} 
\includegraphics[width=0.47\textwidth,height=0.31\textwidth]{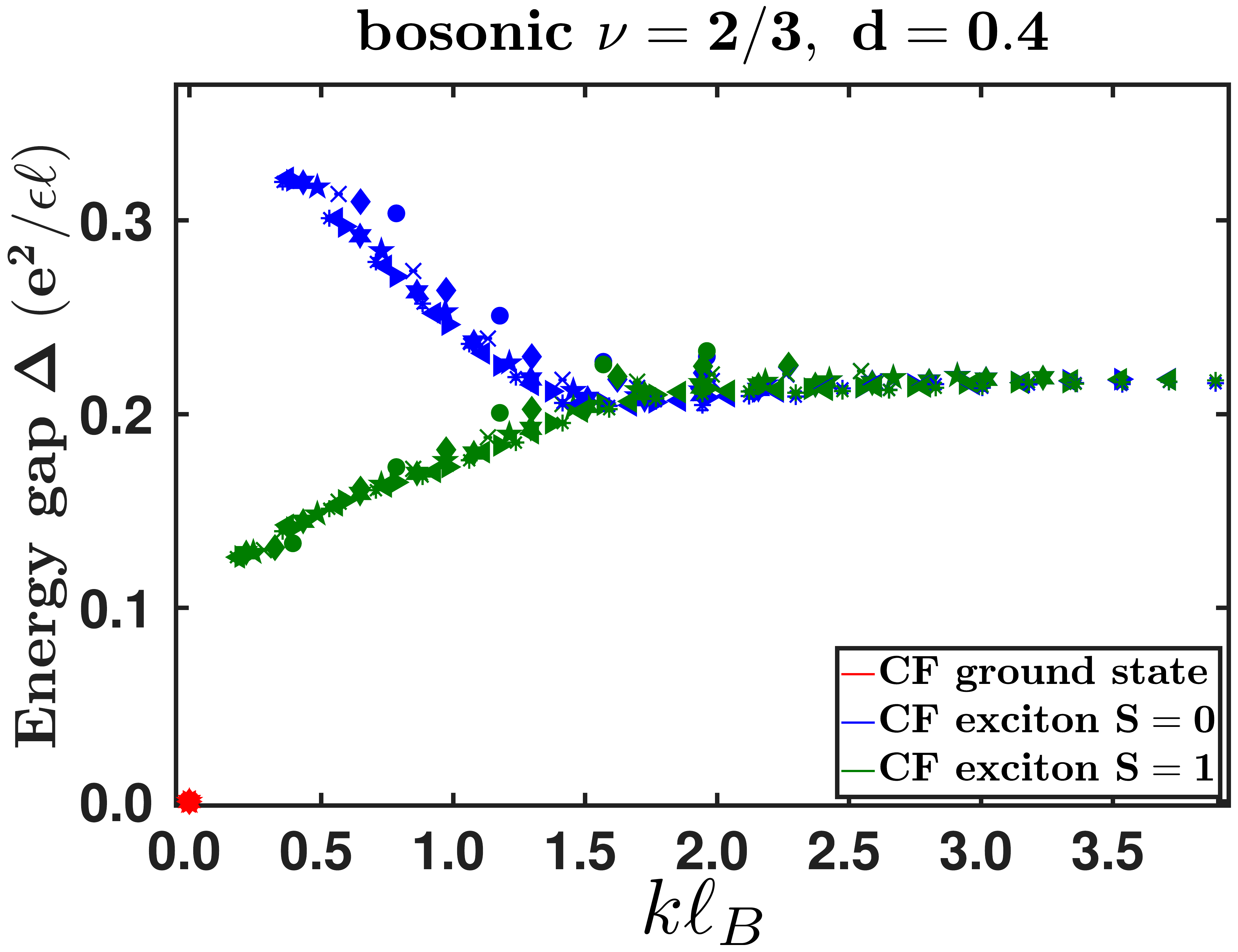} 
\caption{(color online) Dispersion of the collective modes for $N=10$ (dots), $N=18$ (crosses), $N=24$ (pentagrams), $N=30$ (hexagrams), $N=36$ (right-facing triangle), $N=40$ (left-facing triangle) and $N=44$ (asterisk) bosons at $\nu=2/3$ on the sphere at a layer separation of $d/\ell_{B}=0$ (left panel) and $0.4$ (right panel). The $2/3$ Halperin $(221)$ or $\nu=2/3$ bosonic Jain CF pseudospin-singlet state shown in red has total orbital angular momentum $L=0$ and total pseudospin angular momentum $S=0$. The dispersion is plotted as a function of the dimensionless momentum $k\ell_{B}=(L/R)\ell_{B}$, where $R=\sqrt{Q}\ell_{B}$ is the radius of the sphere and $\ell_{B}=\sqrt{\hbar c/(eB)}$ is the magnetic length. The energies are multiplied by $\sqrt{2Q\nu/N}$, the ratio of the density of the finite system to the density in the thermodynamic limit, to alleviate the dependence of the gap on $N$~\cite{Morf86}.}
\label{smfig9}
\end{center}
\end{figure*}

Next, we turn to larger system sizes which are beyond the reach of exact diagonalization. In Fig.~\ref{smfig8} we show the dispersion of the CF modes for several bosonic systems at $\nu=2/3$. We find a nice collapse of the dispersions for different values of $N$ from which we can extract the thermodynamic energies of the two modes in different limits of the wavevector $k\ell_{B}$. For large $k\ell_{B}$ (ideally in the $k\ell_{B}\rightarrow\infty$) we expect the energy of the modes to be independent of $k$ and the dispersions to flatten out. This is because in this limit the constituent CF excitons are far away from each other and do not interact~\cite{Kamilla96b, Kamilla96c, Balram16d}. Therefore, in this regime, we anticipate that the energy of the symmetric and anti-symmetric modes should approach each other. This is precisely what we see in the dispersion of the modes shown in Fig.~\ref{smfig9} with the two modes having an energy of $\Delta\approx 0.24~e^{2}/(\epsilon\ell_{B})$ for $d=0$. For small $k\ell_{B}$ we find the dispersion of the two modes to be quite different from each other with the dipole mode being lower in energy than the quadrupole one. From the plot in Fig.~\ref{smfig9}, we estimate the $k\ell_{B}\rightarrow0$ limit of the dipole and quadrupole mode energies at $d=0$ to be to $\approx0.23~e^{2}/(\epsilon\ell_{B})$ and $\approx0.36~e^{2}/(\epsilon\ell_{B})$ respectively.

\subsection{Collective modes from the single mode approximation (SMA)}
\label{subsec: collective_modes_SMA_221}

The collective modes can also be modeled using the single-mode approximation (SMA)~\cite{Girvin85,Girvin86} where the excitation is created by acting with the density operator $\rho({\bf q})$ on the ground state and projecting the resulting state into the LLL (or by directly acting the ground state with the projected density operator $\bar{\rho}({\bf q})$). In the SMA language, the two low-energy neutral collective modes are obtained by acting on the ground state with the operators
\begin{equation}
\rho^{S}({\bf q}) = \frac{\rho^{\uparrow}({\bf q}) + \rho^{\downarrow}({\bf q})}{\sqrt{2}}, \;\;\;\;\;
\rho^{AS}({\bf q}) = \frac{\rho^{\uparrow}({\bf q}) - \rho^{\downarrow}({\bf q})}{\sqrt{2}}
\end{equation}
where $\rho^{\uparrow}({\bf q})$ and $\rho^{\downarrow}({\bf q})$ are the density operators corresponding to pseudospin $\uparrow$ and $\downarrow$. The quadrupole mode is the obtained from $\rho^{S}$ while the dipole mode is obtained from $\rho^{AS}$. 

\begin{figure*}
	\centerline{\includegraphics[width=\linewidth]{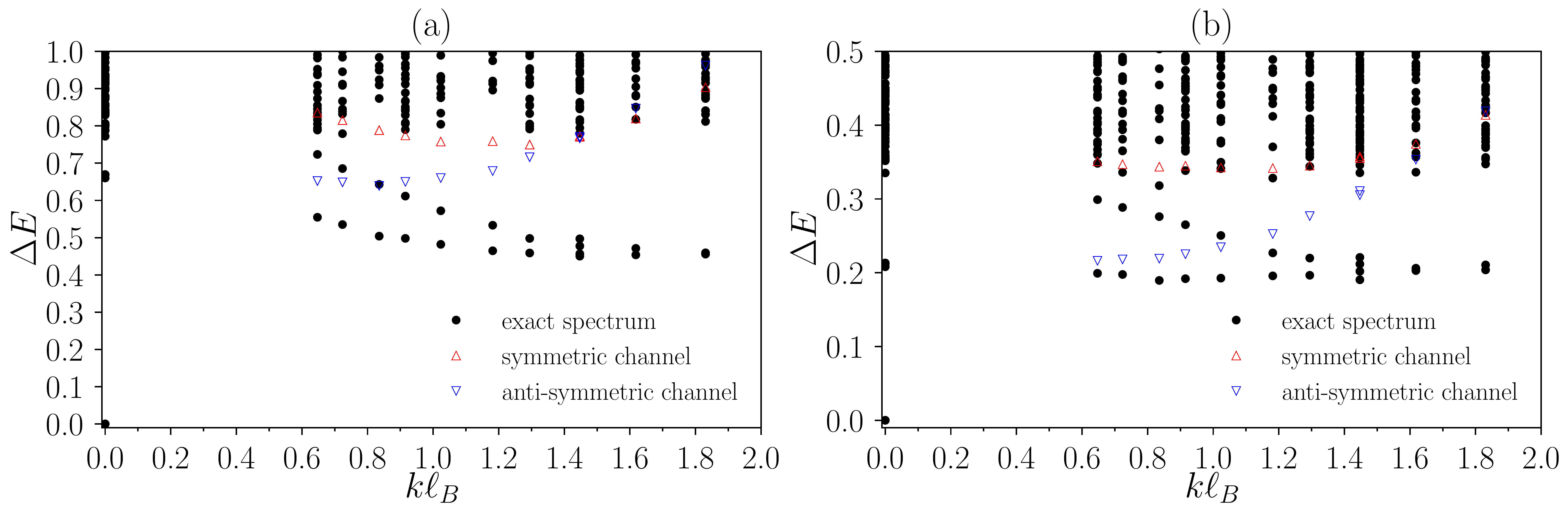}}
	\caption{Comparison of exact spectra with SMA levels in the symmetric and anti-symmetric channels. Here we consider isotropic systems of $N=6,8,$ and $10$ bosons on the torus geometry, interacting via (a) the contact potential and (b) the Coulomb potential with $d=0$. The interlayer displacement ${\bf s}$ is zero. }
	\label{smfig10}
\end{figure*}

How do the SMA modes relate to the exact modes and the ones obtained from the CF theory? In the $k\ell_{B}\rightarrow0$ limit, the SMA mode is exactly equivalent to the excitation obtained from the CF theory~\cite{Kamilla96b, Kamilla96c}. Therefore, in the long-wavelength limit, the SMA modes obtained from $\rho^{S}$ and $\rho^{AS}$ correspond to the symmetric and anti-symmetric CF exciton modes respectively. However, for other values of the momenta $k$, the SMA and CF exciton modes differ from each other. For the LLL Coulomb interaction at small layer separations, the CF theory gives a better description of the collective modes as compared to the one obtained from the SMA~\cite{Kamilla96b, Kamilla96c}. As we see in Fig.~\ref{smfig10}, the SMA modes continue to grow for $k\ell_{B}$ beyond the roton minima and do not flatten out. On the contrary, one expects that for large $k\ell_{B}$ the dispersion flattens out and is independent of $k$ since this regime corresponds to a far-separated quasiparticle-quasihole pair. This effect is not captured by the SMA, as shown in single-layer FQH systems~\cite{Yang12b}.

Another quadrupole mode can be constructed from the excitation containing two composite fermion excitons. For a single component system, the two CF exciton state is known to carry the lowest energy in the $k\ell_{B}\rightarrow0$ limit~\cite{Park00}. For a two-component system, the Hilbert space of the two CF excitons is quite large. Therefore, we have not studied it in this work and defer its investigation in detail to the future.

\section{Bilayer state of bosons at $\nu=2/3$ at large layer separation}

In this section we present evidence to show that the ground state of a single-layer bosonic system at $\nu=1/3$ is a composite Fermi liquid of bosons. This implies that the ground state for the bilayer system of bosons at $\nu=2/3$ at large layer separation, i.e., $d/\ell_{B}\gg 1$, is likely a pair of decoupled CFL of bosons, one in each layer.

To investigate the nature of the ground state of a single-layer bosonic system at $\nu=1/3$, we consider a system of $N$ bosons at the $N_{\phi}=3(N-1)$ flux quanta in the spherical geometry. This flux-particle relationship corresponds to the composite Fermi-liquid state of bosons, where on average the composite bosons (bound state of electrons and an odd number of vortices) see an effective magnetic field which vanishes. The CFL state of bosons at $\nu=1/3$ is described by the wave function:
\begin{equation}
\Psi^{\rm bosonic-CFL}_{1/3} = \mathcal{P}_{\rm LLL} \Phi^{3}_{1} \Phi^{\rm Fermi-sea},
\label{eq_bosonic_CFL_1_3}
\end{equation}
where $\Phi^{\rm Fermi-sea}$ is the Slater determinant wave function of the Fermi-liquid state of electrons. When $N=n^{2}$, the $N$ composite bosons completely fill the lowest $n$ angular momentum shells thereby producing a state which is uniform on the sphere, i.e., has $L=0$. These filled-shell states can be used to construct representatives of the uniform CFL state on the sphere leading to the following wave function~\cite{Rezayi94,Balram15b,Balram17}
\begin{equation}
\Psi^{\rm CFL}_{1/p} = \mathcal{P}_{\rm LLL} \Phi^{p}_{1} \Phi^{\rm filled-shell}~~~~N=4,9,16,25,\cdots,
\label{eq_sphere_CFL_1_p}
\end{equation}
where we have generalized to the $\nu=1/p$ CFL with $p$ odd (even) for bosons (fermions). In Table~\ref{tab:overlaps_CF_filled_shell}, we show the absolute value of the squared overlaps of the exact LLL Coulomb ground state of bosons at $\nu=1/3$ with $\Psi^{\rm CFL}_{1/3}$ [see Eq.~(\ref{eq_sphere_CFL_1_p})] for $N=4$ and $9$ which are the only two systems accessible to exact diagonalization. We find that the overlaps are very close to unity indicating that the ground state of bosons at $\nu=1/3$ is well-represented by the CFL state. For completeness, in Table~\ref{tab:overlaps_CF_filled_shell} we also show the corresponding numbers for fermions at $\nu=1/2$ and $1/4$, where a composite fermion Fermi liquid state has been well-established~\cite{Jain07}. We find that the overlaps of the CFL state at $\nu=1/p$ with the exact LLL Coulomb ground state are comparable for the three values of $p=2,3,4$ considered.

\begin{table}[htb]	
\centering	
\begin{tabular}{|c|c|c||c|c|c||c|c|c|}	
\hline	
\multicolumn{3}{|c|}{$\nu=1/2$ fermions} & \multicolumn{3}{|c|}{$\nu=1/3$ bosons} & \multicolumn{3}{|c|}{$\nu=1/4$ fermions} \\ \hline	
$N$ & $N_{\phi}$ &  $|\langle\Psi^{\rm LLL}|\mathcal{P}_{\rm LLL}\Phi_{1}^{2}\Phi^{\rm filled-shell}\rangle|^{2}$ & $N$ & $N_{\phi}$ &  $|\langle\Psi^{\rm LLL}|\mathcal{P}_{\rm LLL}\Phi_{1}^{3}\Phi^{\rm filled-shell}\rangle|^{2}$ & $N$ & $N_{\phi}$ & $|\langle\Psi^{\rm LLL}|\mathcal{P}_{\rm LLL}\Phi_{1}^{4}\Phi^{\rm filled-shell}\rangle|^{2}$\\ \hline	
4 &  6  & 1.0000~\cite{Wu93} & 4 &  9 & 1.0000 & 4 & 12 & 0.9999~\cite{Balram20a} \\ \hline 	
9 & 16 & 0.9988~\cite{Yang19a} & 9 & 24 & 0.9955 & 9 & 32 & 0.9845 \\ \hline 	
\end{tabular}	
\caption{\label{tab:overlaps_CF_filled_shell} Absolute value of the squared overlaps of the filled-shell composite Fermi liquid states $|\mathcal{P}_{\rm LLL}\Phi_{1}^{p}\Phi^{\rm filled-shell}\rangle$ [see Eq.~(\ref{eq_sphere_CFL_1_p})] at $\nu=1/p$ for $p=2,3,4$ with the exact lowest Landau level Coulomb ground state $|\Psi^{\rm LLL}\rangle$ for a single layer system of $N$ particles at $N_{\phi}=p(N-1)$ flux quanta in the spherical geometry. }	
\end{table}

\begin{figure*}[tbhp]		
    \centerline{\includegraphics[width=0.33\linewidth]{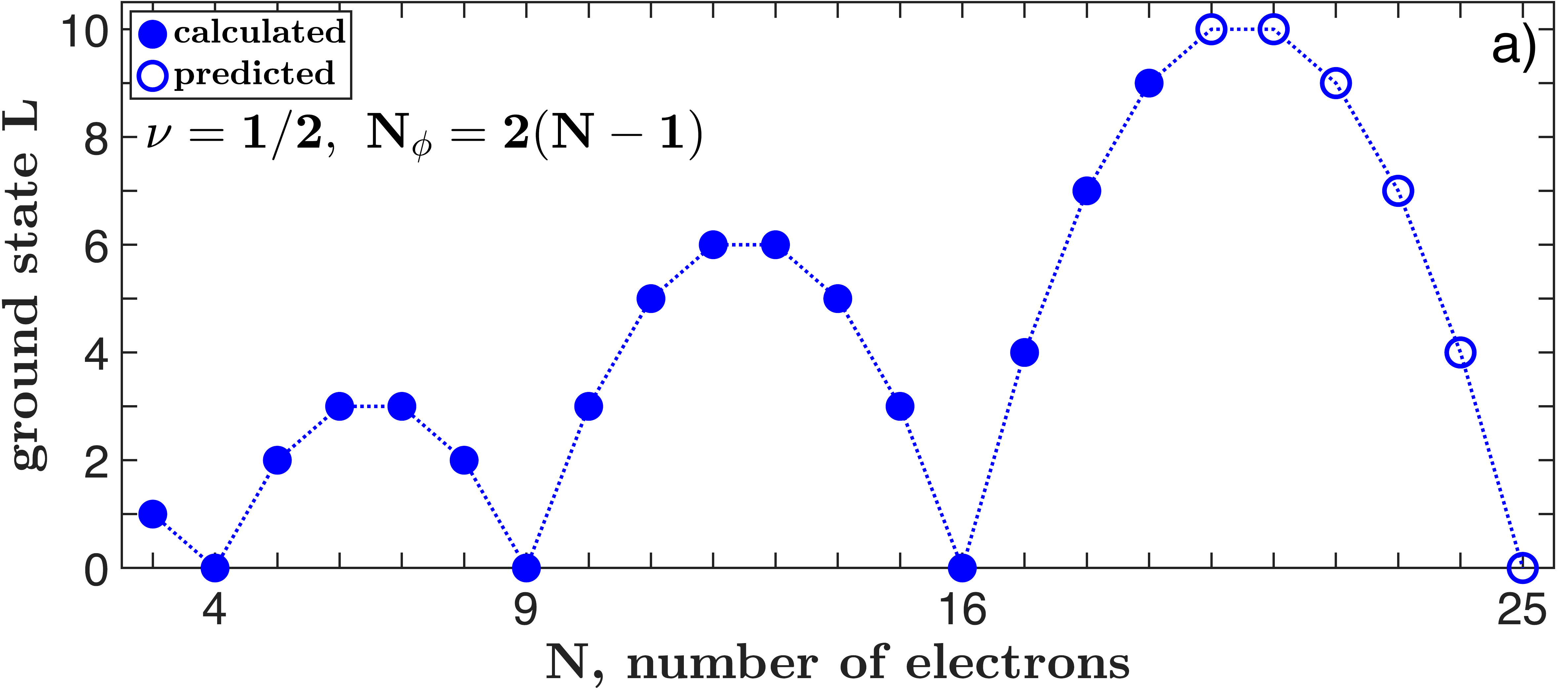}
	\includegraphics[width=0.33\linewidth]{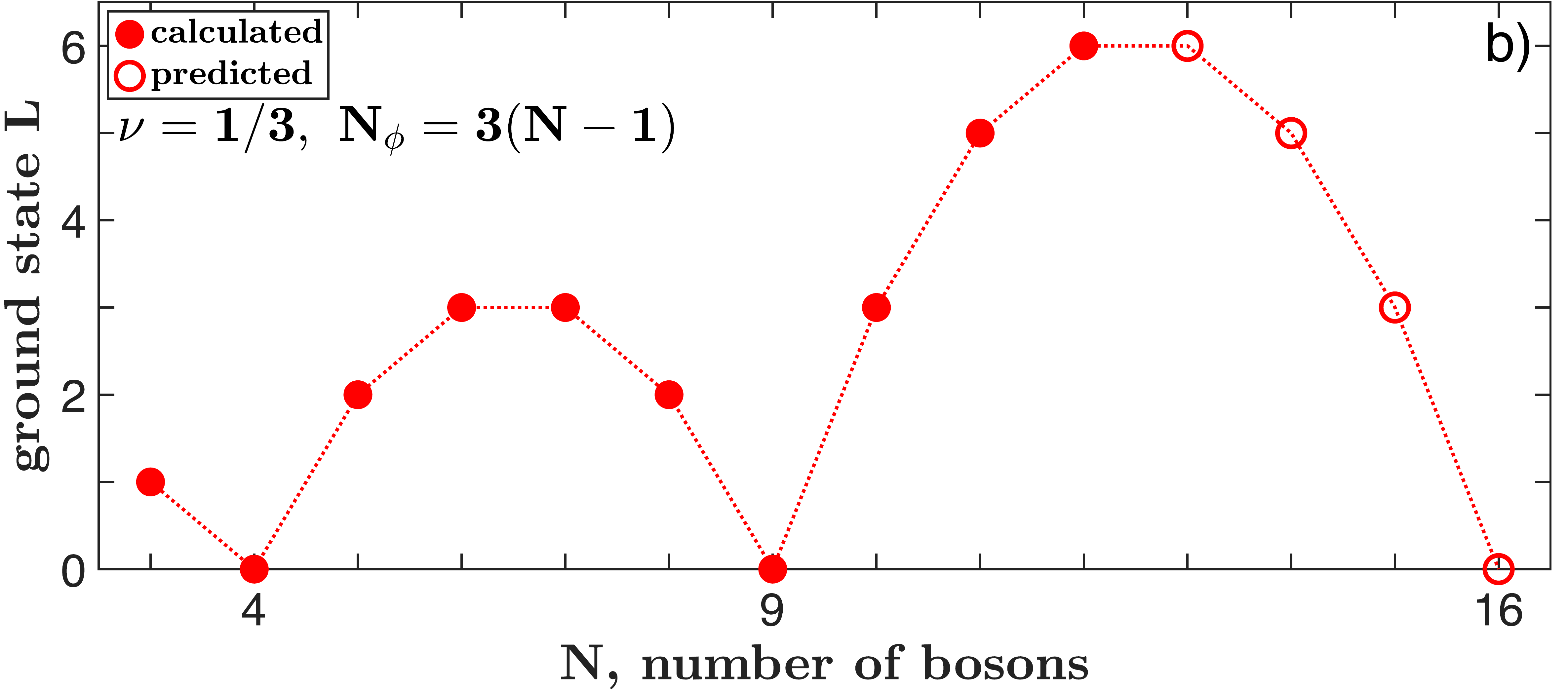}
    \includegraphics[width=0.33\linewidth]{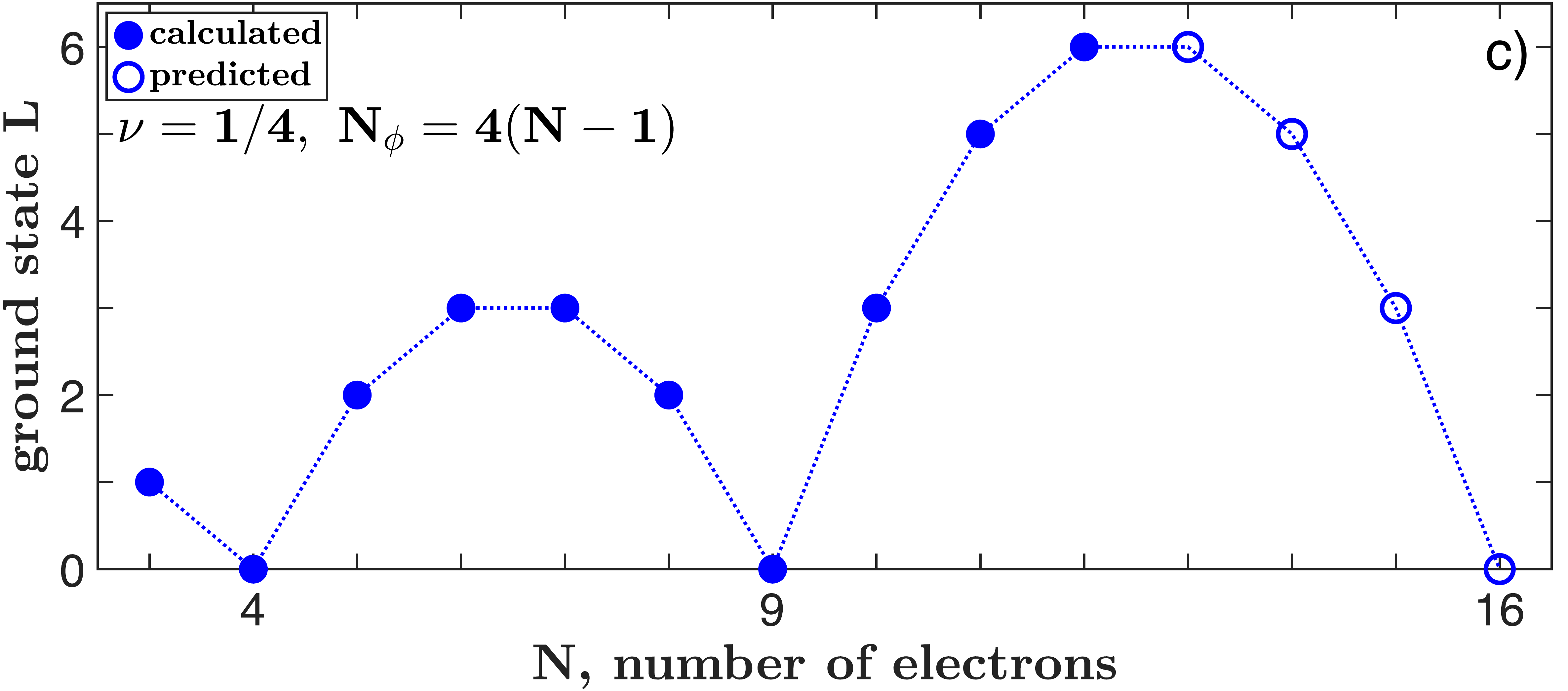}}
	\caption{The total orbital angular momentum $L$ of the lowest Landau level Coulomb ground state at $N_{\phi}=3(N-1)$ as a function of particle number $N$ obtained from exact diagonalization in the spherical geometry for bosons at $\nu=1/3$ [center panel b)].
For comparison, we also show the analogous plot for the fermionic systems at $N_{\phi}=p(N-1)$ for a) $p=2$, $\nu=1/2$ (left panel) and c) $p=4$, $\nu=1/4$ (right panel). Filled symbols represent calculated values while open symbols are predictions based on the Hund rule (see text).}
	\label{smfig11}
\end{figure*}

To further corroborate our interpretation of the $1/3$ state of bosons as a CFL, we consider the exact LLL Coulomb ground state of $N$ bosons at $N_{\phi}=3(N-1)$ at values of $N\neq n^{2}$. When $N\neq n^{2}$, an angular momentum shell will be partially occupied which would generically result in a non-uniform ground state with $L>0$. In Fig.~\ref{smfig11}(b) we show the total orbital angular momentum $L$ of the ground state obtained from the exact diagonalization of the Coulomb interaction in the LLL of a system of $N$ bosons at $3(N-1)$ flux quanta. Following the discussion in the previous paragraph as expected we find a uniform (with $L=0$) ground state for $N=4,9$. Interestingly, for other values of $N$ the total orbital angular momentum is the \emph{maximum} value that can be obtained by combining the angular momenta of the quasiparticles or quasiholes in the topmost partially filled shell. This result is reminiscent of Hund's rule in atomic physics whereby the large values of $L$ allow the particles to avoid each other maximally resulting in minimizing their interaction energy. Analogous calculations for the $1/2$ state of fermions were first carried out by Rezayi and Read~\cite{Rezayi94}. Since we find identical values of $L$ for the $1/3$ state of bosons as for the $1/2$ state of fermions we conclude that the former (just like the latter) should be described as a CFL. For completeness, we have expanded on the Rezayi and Read calculation considering larger systems at $1/2$ and also fermionic states at $1/4$. These results are shown in Figs.~\ref{smfig11}(a) and \ref{smfig11}(c) and are fully consistent with the Hund rule.  

\section{Interaction of spin-$1$ modes}

Here we briefly discuss the pairing of the spin-$1$ modes from the point of view of the effective theory. First, we observe that the spin-$1$ mode itself does not respond to the geometric quench. Indeed the coupling of the spin-$1$ mode directly to the spin-$2$ mode takes the form of a modified mass term
\begin{equation}
\mathcal L =  -\epsilon^{ij} v_i \dot{v}_j - M  g^{ij} v_i v_j\,.
\end{equation}
The equations of motion suggest that $v_i=0$ remains the solution after switching on the metric field.

In order to get a non-trivial dynamics we have to assume that the spin-$1$ modes interact with each other. We consider the following interacting Lagrangian
\begin{equation}
\mathcal L =  -\epsilon^{ij} v_i \dot{v}_j - M  |v|^2 - \lambda \delta^{ij} \delta^{kl} v_i v_j v_k v_l.
\end{equation}
The last term can be decoupled with a Hubbard-Stratonovich transformation
\begin{equation}
-\lambda \delta^{ij} \delta^{kl} v_i v_j v_k v_l \quad \longrightarrow \quad \lambda^{-1} h_{ij} h_{kl} \delta^{ij} \delta^{kl} - 2 h^{ij} v_i v_j\,.
\end{equation}
The equations of motion 
\begin{equation}
h_{ij} = \lambda v_i v_j\,,
\end{equation}
is degenerate, however $\delta_{ij} + h_{ij}$ is a proper spin-$2$ field. The new Lagrangian takes form
\begin{equation}
\mathcal L = -\epsilon^{ij} v_i \dot{v}_j - M \left(\delta^{ij} + \frac{2}{M}h^{ij}\right) v_i v_j + \lambda^{-1} h_{ij} h_{kl} \delta^{ij} \delta^{kl}\,.
\end{equation}
Integrating out the gapped field $v_i$ is possible technically, but is not strictly legal since there is no scale separation between the masses of $v_i$ and $h_{ij}$. It is, however possible to imagine running renormalization group (RG) for $v_i$. The RG will generate all terms for $h_{ij}$ allowed by symmetries. In that case the low energy Lagrangian for $h_{ij}$ will take form 
\begin{equation}
\mathcal L = \alpha \epsilon^{ij} h_{ik} \dot{h}_{kj} + 2\beta  h_{ij} \delta^{ij} + \ldots\,,
\end{equation}
where, $\alpha$ and $\beta$ are unknown constants and $\beta/\alpha \approx 2M$ determines the gap of the emergent spin-$2$ mode.

\end{document}